\documentclass[12pt,showpacs,showkeys,amsmath,amssymb]{revtex4}
\usepackage{amsmath,amsfonts,amsthm,amscd,amssymb,latexsym}
\usepackage{bm}
\usepackage{dcolumn}
\usepackage{graphicx}
\usepackage{epstopdf}
\usepackage{color}
\usepackage{epsf}
\usepackage{epsfig}
\usepackage{graphicx, epic, eepic, color}

\newcommand{\beq}{\begin{equation}}
\newcommand{\eeq}{\end{equation}}

\begin{document}

\title{Twisted Gravitational Waves in the Presence of a Cosmological Constant}

\author{Hassan \surname{Firouzjahi}$^{1}$}
\email{firouz@ipm.ir}
\author{Bahram \surname{Mashhoon}$^{1,2}$}
\email{mashhoonb@missouri.edu}

\affiliation{$^1$School of Astronomy, Institute for Research in Fundamental
Sciences (IPM), P. O. Box 19395-5531, Tehran, Iran\\
$^2$Department of Physics and Astronomy, University of Missouri, Columbia, Missouri 65211, USA\\
}

\date{\today}

\begin{abstract}
We find exact nonlinear solutions of general relativity that represent twisted gravitational waves (TGWs) in the presence of a cosmological constant. A TGW  is a  nonplanar wave propagating along a fixed spatial direction with a null Killing wave vector that has a nonzero twist tensor. The solutions all turn out to have wave fronts with negative Gaussian curvature. Among the classes of solutions presented in this paper, we find a unique class of simple  conformally flat TGWs that is due to the presence of a negative cosmological constant and therefore represents part of anti-de Sitter spacetime. The properties of this special solution are studied in detail. 
\end{abstract}


\maketitle

\section{Introduction}

A twisted gravitational wave is a free nonlinear unidirectional radiative solution of general relativity (GR) such that its null propagation vector $k$ has a nonzero twist tensor~\cite{Bini:2018gbq, Bini:2018iyu, Rosquist:2018ore}. Imagine a gravitational field described by the metric $ds^2 = g_{\mu \nu} (t-z, x, y)\, dx^\mu dx^\nu$ in $x^\mu = (t, x, y, z)$ coordinates that represents a gravitational wave propagating along the $z$ direction. Throughout this paper, we use units such that $c = G = 1$; moreover, the signature of the spacetime metric is +2 and greek indices run from 0 to 3, while latin indices run from 1 to 3. Let $k= \partial_t + \partial_z$ be the null propagation vector of the wave; then, 
\beq \label{I1}
k_\mu\,k^\mu = 0\,, \qquad  k_{\mu ; \nu} + k_{\nu ; \mu} = 0\,.
\eeq
It follows from these relations that $k_{\mu ; \nu}\,k^{\nu} = 0$, so that the spacetime under consideration admits a null geodesic Killing vector field that is nonexpanding and shearfree. 

Spacetimes that admit a covariantly constant null vector field $k$ with $k_\mu\,k^\mu = 0$ and $k_{\mu ; \nu} = 0$ represent \emph{plane-fronted gravitational waves with parallel rays} ($pp$-waves). These were first discovered in 1925 by Brinkmann~\cite{Brink} and have since been the subject of detailed investigations~\cite{BPR, Khan:1971vh, JBG, R1, Griffiths:2009dfa}. As discussed in Ref.~\cite{Bini:2018gbq}, \emph{plane gravitational waves} have at least five Killing vector fields and form a subclass of $pp$-waves. From
\beq \label{I2}
k_{\mu ; \nu} = \frac{1}{2}\,(k_{\mu ; \nu} + k_{\nu ; \mu}) + \frac{1}{2}\,(k_{\mu ; \nu} - k_{\nu ; \mu})\,, 
\eeq   
we note that if the twist tensor 
\beq \label{I3}
 \mathbb{T}_{\mu \nu} =  \frac{1}{2}\,(k_{\mu ; \nu} - k_{\nu ; \mu}) = k_{[\mu , \nu]}\,
\eeq
of the gravitational wave under discussion in Eq.~\eqref{I1} vanishes, then our assumptions in Eq.~\eqref{I1} lead to $k_{\mu ; \nu} = 0$ and hence we have a $pp$-wave. On the other hand, if $\mathbb{T}_{\mu \nu}\ne 0$, we then have twisted waves that are \emph{nonplanar}; that is, they have nonuniform wave fronts with nonzero Gaussian curvature. 

To illustrate these ideas in more detail, let us assume a spacetime metric of the form
\begin{equation}\label{I4}
ds^2 = -\gamma_0\, (dt^2 -dz^2)+ \gamma_1\,dx^2+ 2\gamma_2\, dx\,dy + \gamma_3\,dy^2\,,
\end{equation}  
where $\gamma_\mu = \gamma_\mu (t-z, x, y)$. The $(t, x, y, z)$ coordinate system is physically admissible~\cite{Bini:2012ht} if $\gamma_0 > 0$, $\gamma_1 > 0$, $\gamma_3 > 0$ and $\Delta := \gamma_1\,\gamma_3 - \gamma_2^2 > 0$. In these coordinates, $k^\mu = (1, 0, 0, 1)$ is the null propagation Killing vector field. It is useful to introduce the retarded and advanced null coordinates $u = t-z$ and $v = t + z$, respectively, and write metric~\eqref{I4} as  
\begin{equation}\label{I5}
ds^2 = -\gamma_0\, du\,dv+ \gamma_1\,dx^2+ 2\gamma_2\, dx\,dy + \gamma_3\,dy^2\,,
\end{equation} 
where $k = 2\,\partial_v$. The wave front corresponds to hypersurfaces of constant $u = u_0$, in which case the metric reduces to 
\begin{equation}\label{I6}
d\sigma^2 = ds^2\big|_{u = u_0} = \gamma_1(u_0, x, y)\,dx^2+ 2\gamma_2(u_0, x, y)\, dx\,dy + \gamma_3(u_0, x, y)\,dy^2\,.
\end{equation}
The Gaussian curvature of this surface vanishes for $pp$-waves and is nonzero for twisted gravitational waves (TGWs). The formula for the Gaussian curvature is given in Appendix A. For metric~\eqref{I4}, the null propagation vector $k$ is normal to the wave front, namely, 
\begin{equation}\label{I7}
k_\mu = \gamma_0(u, x, y)\,(-1, 0, 0, 1) = -\gamma_0(u, x, y)\, \frac{\partial u}{\partial x^\mu}\,.
\end{equation}
It follows that in this case, the wave's twist tensor~\eqref{I3} is given by
\begin{equation}\label{I8}
\mathbb{T}_{\mu \nu} = \frac{1}{2} \left(\frac{\partial\gamma_0}{\partial x^\mu}\,\frac{\partial u}{\partial x^\nu} - \frac{\partial\gamma_0}{\partial x^\nu}\,\frac{\partial u}{\partial x^\mu}\right)\,.
\end{equation}
If $\gamma_0$ is only a function of $u$, or if it is a constant independent of coordinates, then $\mathbb{T}_{\mu \nu} = 0$ and we have a $pp$-wave; otherwise, $\mathbb{T}_{\mu \nu} \ne 0$ and we have a TGW. Using Eq.~\eqref{I7}, Eq.~\eqref{I8} can be written as
\begin{equation}\label{I9}
\mathbb{T}_{\mu \nu} = \frac{1}{2\gamma_0} \left(\frac{\partial\gamma_0}{\partial x^\nu}\,k_\mu - \frac{\partial\gamma_0}{\partial x^\mu}\,k_\nu\right)\,,
\end{equation}
where 
\begin{equation}\label{I10}
\frac{\partial\gamma_0}{\partial x^\mu}\,k^\mu = 0\,,
\end{equation}
since $2\,\partial_v \gamma_0 =  \gamma_{0, t} +  \gamma_{0, z} = 0$.
Equations~\eqref{I9} and~\eqref{I10} imply that the \emph{twist scalar} $\omega$ vanishes in this case
\begin{equation}\label{I11}
\omega^2 := \frac{1}{2}\,\mathbb{T}_{\mu \nu} \mathbb{T}^{\mu \nu} = 0\,.
\end{equation}
This result is in agreement with the theorem that $\omega = 0$ if and only if the null geodesic congruence is hypersurface-orthogonal.

The TGWs under consideration here belong to the Kundt class of solutions of GR~\cite{R1, Griffiths:2009dfa}. As demonstrated in Appendix A of Ref.~\cite{Bini:2018iyu}, it is possible to write our TGW metric~\eqref{I5} in Kundt's form. The Kundt solutions have been extensively studied and the solutions presented in this paper are probably known in some form in other coordinate systems.

No reasonable astronomical source of TGWs is known. Gravitational radiation emitted by known astronomical sources are expected to have expanding nearly spherical wave fronts far from the source. Therefore, TGWs have been tentatively interpreted in terms of running cosmological waves~\cite{Bini:2018gbq, Bini:2018iyu, Rosquist:2018ore}. Observations of distant supernovae have led to the discovery of the accelerating expansion of the universe. The standard cosmological models that take this acceleration into account involve a positive cosmological constant $\Lambda$.
Thus we might expect that TGWs should be  compatible with the existence of a cosmological constant. However, in previous work on TGWs~\cite{Bini:2018gbq, Bini:2018iyu, Rosquist:2018ore}, the cosmological constant was set equal to zero in order to make the field equations tractable.  It is therefore important to look for TGWs in the presence of a nonzero cosmological constant. This issue will be addressed in the present paper.   We seek TGW solutions of the gravitational field equations in vacuum but with a cosmological constant $\Lambda$, namely, 
\beq \label{I12}
R_{\mu \nu} = \Lambda\, g_{\mu \nu}\,,
\eeq
where $R_{\mu \nu} := R^{\alpha}{}_{\mu \alpha \nu}$ is the Ricci tensor. To render the resulting differential equations manageable,  we assume a solution of the form~\eqref{I5} such that the gravitational potentials are all functions of the dimensionless variable
\beq \label{I13}
w = s\, u + p\, x + q\, y\,,
\eeq
where $s$, $p$ and $q$ are in general nonzero constant parameters of dimensions 1/length. We must specifically assume that $s \ne 0$ throughout; otherwise, the solution is static and cannot represent a wave. Moreover, it is clear that by a simple coordinate translation we can add any constant to $w$; henceforth, such constants will be ignored throughout with no loss in generality. There was a preliminary indication in previous work that such solutions may accommodate a cosmological constant~\cite{Bini}. We assume throughout that $\gamma_0 ' := d\gamma_0/dw \ne 0$; otherwise, $\mathbb{T}_{\mu \nu} = 0$ and the solution would represent $pp$-waves. The gravitational field equations~\eqref{I12} are worked out explicitly in Appendix B for metric~\eqref{I5} when the metric coefficients are all functions of $w$ defined in Eq.~\eqref{I13}. 

The Riemann curvature tensor can be decomposed into its Weyl, Ricci and scalar curvature components. Therefore, the Kretschmann scalar $\mathcal{K}$ can be expressed in general as
\beq \label{I14}
\mathcal{K} := R_{\alpha \beta \gamma \delta}\,R^{\alpha \beta \gamma \delta} = \mathcal{W} + 2\,R_{\mu \nu}\,R^{\mu \nu} -\frac{1}{3}\,R^2\,, \qquad \mathcal{W} := C_{\alpha \beta \gamma \delta}\,C^{\alpha \beta \gamma \delta}\,,
\eeq
where $C_{\alpha \beta \gamma \delta}$ is the Weyl conformal curvature tensor and $R$ is the scalar curvature. For the solutions of GR that satisfy Eq.~\eqref{I12}, $R = 4\,\Lambda$ and
\beq \label{I15}
\mathcal{K} = \frac{8}{3} \Lambda^2 + \mathcal{W}\,.
\eeq
The TGWs that we study in this paper are such that $\mathcal{W} \ge 0$ and $\mathcal{W} \propto \gamma_0^{-3}$. For $\mathcal{W} > 0$, our solutions are singular at $\gamma_0 = 0$ where $\mathcal{W}$ diverges.

In Section II, we present the general solution of the field equations~\eqref{I12} for metric~\eqref{I4} where the gravitational potentials are only functions of $w$; furthermore, we assume that there is no cross term ($\gamma_2 = 0$) and $q = 0$ in Eq.~\eqref{I13} for the sake of simplicity. These latter restrictions are in turn removed in Sections III and IV, respectively. That is, we keep $q = 0$, but extend our results to the case where a cross term is present ($\gamma_2 \ne 0$) in Section III and in Section IV, we return to the setting of Section II with no cross term, but with $q \ne 0$  in Eq.~\eqref{I13}.   In the presence of both positive and negative $\Lambda$, we find classes of TGW solutions in Sections II - IV. Each TGW depends on the solution of an ordinary differential equation for $A(w) :=\ln{\gamma_0}$. We study the general character of these TGW solutions with $\Lambda \ne 0$; moreover, we determine their curvature singularities and the algebraic properties of their Weyl curvature tensors within the Petrov classification scheme. From the results of Sections II and IV, a simple unique conformally flat TGW solution is found for negative $\Lambda$ such that $g_{\mu \nu} = w^{-2}\, \eta_{\mu \nu}$ and $p^2 + q^2 = -\Lambda /3$.  It turns out to be part of the anti-de Sitter spacetime manifold. Various properties of this solution are investigated in detail in Sections V and VI. A discussion of our results is contained in Section VII.

\section{TGWs in the presence of $\Lambda$}

Consider a metric of the form
\begin{equation}\label{N1}
 ds^2 = - e^{A(w)}(dt^2-dz^2) + e^{B(w)}\,dx^2 + e^{C(w)}\,dy^2\,,
\end{equation}
where 
\begin{equation}\label{N2}
 w = s\,u + p\,x\,, \qquad s \ne 0\,, \qquad p \ne 0\,
\end{equation}
and $u = t - z$ is the retarded null coordinate. This spacetime contains three Killing vector fields, namely, $\partial_t + \partial_z$,  $p\,\partial_t -s\, \partial_x$ and $\partial_y$. The gravitational field equations in this case reduce to the following five equations:
\beq \label{N3}
p^2\,A'(A'+2\,C')+4\,\Lambda \,e^B=0\,,
\eeq
where $A':= dA/dw$, etc., 
\beq \label{N4}
2\,A'' +A'^2 = A'(B'+C')\,,
\eeq
\beq \label{N5}
2\,C'' +C'^2  = A'^2 + B'\,C'\,,
\eeq
\beq \label{N6}
2\,A''  + 2\,C'' + C'^2 = A'(B'+C') + B'C'\,,
\eeq
\beq \label{N7}
2\,A' (B'+C')  = B'^2+C'^2 + 2\,B''+ 2\,C''\,,
\eeq
cf. Appendix B. These results  can also be obtained from the field equations in the presence of $\Lambda$ given in Appendix A of Ref.~\cite{Bini:2018gbq}.

Equation~\eqref{N6} is equivalent to the sum of Eqs.~\eqref{N4} and~\eqref{N5}. After dividing both sides of Eq.~\eqref{N4} by $A' \ne 0$, the resulting equation can be simply integrated and we get
\beq \label{N8}
A' = 2\,k_A\, e^{\frac{1}{2}\,(B+C-A)}\,,
\eeq
where $k_A \ne 0$ is a dimensionless integration constant. Furthermore, the sum of Eqs.~\eqref{N6} and~\eqref{N7} can be integrated once and the result is
\beq \label{N9}
B' - A' = k_B\, e^{-\frac{1}{2}\,(B+C)}\,;
\eeq
similarly, the difference between Eqs.~\eqref{N4} and~\eqref{N5}  can also be integrated and we find
\beq \label{N10}
A' - C' = k_C\, e^{-\frac{1}{2}\,(2\,A - B+C)}\,,
\eeq
where $k_B$ and $k_C$ are dimensionless constants of integration. 

Let us now start with Eq.~\eqref{N8} and define a function $F(w)$,
\beq \label{N11}
F := e^{\frac{1}{2}\,A}\,, \qquad F' = k_A\, e^{\frac{1}{2}\,(B+C)}\, \qquad A' = 2\,\frac{F'}{F}\,,
\eeq
in terms of which $B'$ and $C'$ can be written using Eqs.~\eqref{N9} and~\eqref{N10} as 
\beq \label{N12}
B'  = 2\,\frac{F'}{F}+ \frac{k_A\,k_B}{F'}\,, \qquad C'  = 2\,\frac{F'}{F} - \frac{k_C}{F^2}\,e^{\frac{1}{2}\,(B-C)}\,.
\eeq
We note that for $A\in (-\infty, \infty)$, $F \in (0, \infty)$. To calculate $C'$ in terms of $F$, we go back to Eq.~\eqref{N4} and find
\beq \label{N13}
A'\,C'  = 2\,\left(2\,\frac{F''}{F} - 2\,\frac{F'^2}{F^2} - \frac{k_A\,k_B}{F}\right)\,.
\eeq
Substituting this result in Eq.~\eqref{N3}, we get
\beq \label{N14}
2\,\frac{F''}{F} - \frac{F'^2}{F^2} - \frac{k_A\,k_B}{F}+\lambda \,e^B=0\,,
\eeq
where 
\beq \label{N15}
\lambda := \frac{\Lambda}{p^2}\,
\eeq
is in this case the dimensionless \emph{reduced cosmological constant}. 

Let us now return to Eq.~\eqref{N11} and note that 
\beq \label{N16}
F'' = \frac{1}{2}\, k_A\, (B' +C')\,e^{\frac{1}{2}\,(B+C)}\,.
\eeq
Next, using Eq.~\eqref{N12} we find
\beq \label{N17}
B' + C' = 4\,\frac{F'}{F}+ \frac{k_A\,k_B}{F'} - \frac{k_C}{F^2}\,e^{\frac{1}{2}\,(B-C)}\,.
\eeq
Substituting this relation in Eq.~\eqref{N16} results in 
\beq \label{N18}
2\,F'' - 4\,\frac{F'^2}{F} - k_A\,k_B = - \frac{k_A\,k_C}{F^2}\,e^B\,,
\eeq
where Eq.~\eqref{N11} has been employed as well. Assuming that $k_C \ne 0$, we can find $\exp{(B)}$ from Eq.~\eqref{N18} and substitute it in Eq.~\eqref{N14} to find an autonomous second order ordinary differential equation for the function $F(w)$. Indeed, we get
\beq \label{N19}
F\,(2\,F''-  \beta)\,(1+\alpha F^3)- F'^2\,(1+4\,\alpha\,F^3) = 0\,,
\eeq
where
\beq \label{N20}
\alpha := - \frac{\lambda}{k_A\,k_C}\,, \qquad \beta := k_A\,k_B\,.
\eeq
The first integral of Eq.~\eqref{N19} can be determined by writing $2\,F'' =  d(F'^2)/dF$ and integrating the resulting equation for $F'^2$. We find 
\beq \label{N21}
F'^2 + \mathbb{V}(F) = 0 \,, \qquad \mathbb{V}(F) = - F\,(1+\alpha\, F^3)\left[ \kappa - \frac{1}{3}\,\beta \ln(\alpha + F^{-3}) \right]\,,
\eeq
where $\kappa$ is a new dimensionless integration constant. In terms of $F > 0$, the metric functions are
\beq \label{N22}
e^A = F^2\,, \quad e^B = \frac{F'^2}{k_A\,k_C}\,\left(\frac{3F}{1+\alpha\, F^3}\right)\,, \quad e^C = \frac{k_C}{k_A}\,\left(\frac{3F}{1+\alpha\, F^3}\right)^{-1}\,.
\eeq
These general solutions have Weyl curvature tensors that are algebraically special and of type II in the Petrov classification.  The Kretschmann scalar, $\mathcal{K} := R_{\mu \nu \rho \sigma}\,R^{\mu \nu \rho \sigma}$, for this class of solutions is positive and is given by
\beq \label{N22a}
\mathcal{K}_{II} = \frac{8}{3}\, \Lambda^2 + \frac{4}{3}\,\frac{k_A^2\,k_C^2\,p^4}{F^6(w)}\,,
\eeq
so that we have a curvature singularity at $F = 0$, as expected. 

To gain insight into the nature of this class of solutions, we note that Eq.~\eqref{N21} can be interpreted in terms of one-dimensional motion of a classical particle with zero total energy in the effective potential $\mathbb{V}$. In this paper, we employ the effective potential energy method to characterize the nature of the TGW solutions with a cosmological constant. In the plots of the effective potentials in the cases we consider, the amount of available kinetic energy  is given by the absolute magnitude of the effective potential below the horizontal axis.  In this way, it is possible to give a qualitative description of the behavior of $F$ as a function of $w$. 

\begin{figure}
\includegraphics[scale=0.41]{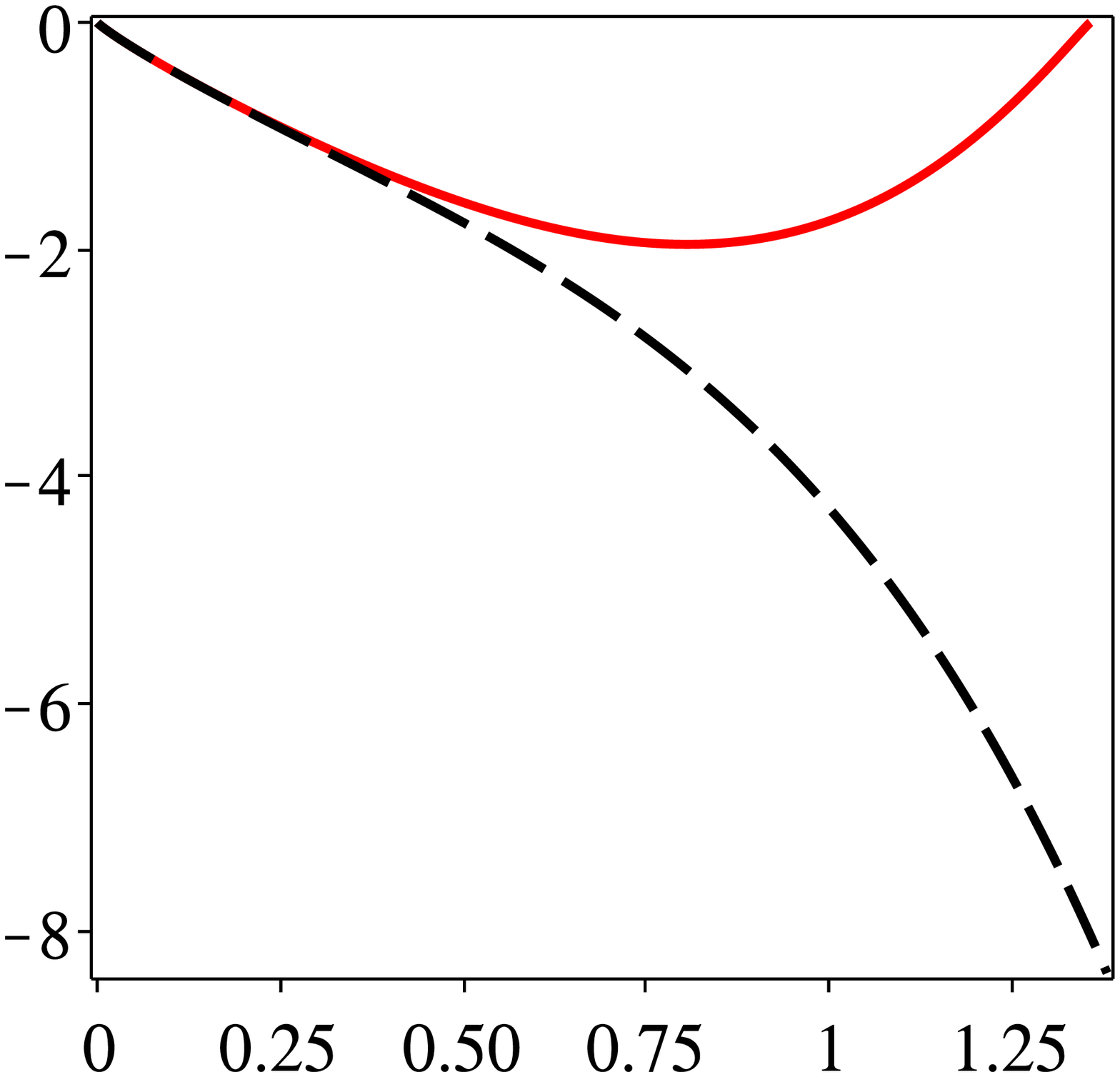} \hspace{0.3cm}
\includegraphics[scale=0.4]{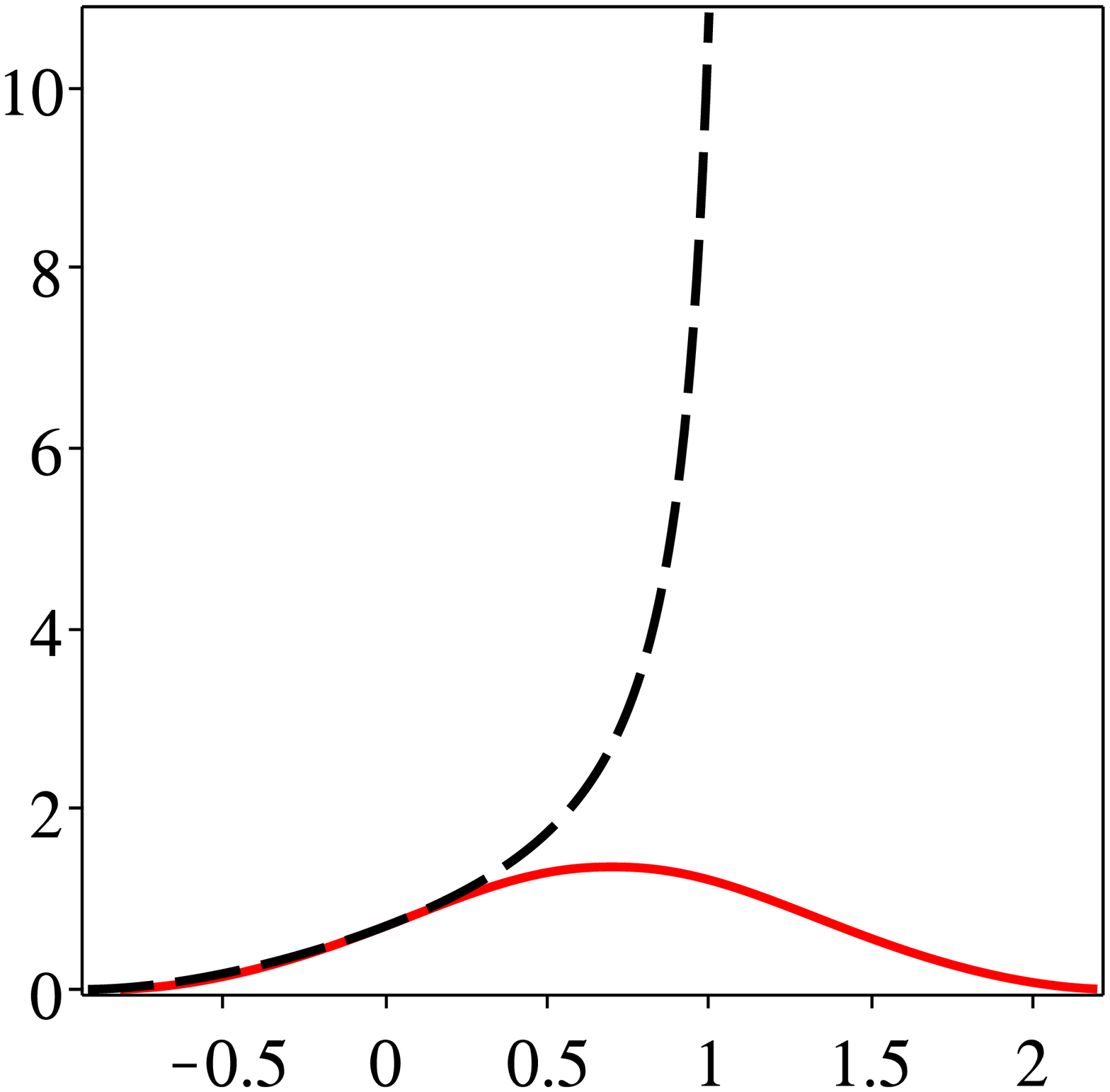}
\caption{\label{figII} Left panel: Plot of the effective potential $\mathbb{V}$ versus $F$ for zero total energy and parameter values $\beta = -0.5$, $\kappa = 3$ and $\alpha = \pm\, 0.4$. In fact,  
$\alpha = 0.4$ for $\Lambda <0$, while $\alpha = -0.4$ for  $\Lambda >0$. The solid red curve is for $\Lambda>0$, while the dashed black curve is for $\Lambda <0$.  Right panel: Plot of $F(w)$ versus $w$.  We integrate Eq.~\eqref{N21} for $F(w) \ge 0$ and $F(0) = 0.7$ with the same parameter values as in the left panel. The solid red curve is for $\Lambda>0$, while the dashed black curve is for $\Lambda<0$. In the latter case, $F(w)$ diverges at $w \approx 1.1$.}
\end{figure}

We have plotted the effective potential $\mathbb{V}$ in the left panel of Figure~\ref{figII} for both signs of the cosmological constant. In our numerical work, we assumed that $k_A > 0$, $k_B < 0$ and $k_C > 0$; more specifically, we choose $\alpha = \pm\, 0.4$, $\beta = - 0.5$ and $\kappa = 3$. For $\Lambda > 0$, $\alpha < 0$ and $\mathbb{V}$ vanishes at $F = 0$ and $F = (-\alpha)^{-1/3}$, see the solid red curve in the left panel of Figure~\ref{figII}.  Between these turning points, the motion is oscillatory and we have plotted $F(w)$ for a complete period of this oscillation in the right panel of Figure~\ref{figII}, see the solid red curve in the interval from $w \approx -0.84$ to $w \approx 2.24$.  For $\Lambda < 0$, $\alpha > 0$ and $\mathbb{V}$ vanishes at $F = 0$, monotonically decreases with increasing $F$  and diverges as $F \to \infty$, see the dashed black curve in the left panel of Figure~\ref{figII}. In this case, $F(w)$ vanishes at $w \approx - 0.92$ and diverges at the endpoints of a \emph{finite} interval in $w$ from $\approx - 2.94$ to $\approx 1.1$. Only half of this interval is plotted in the right panel of Figure~\ref{figII}; in fact, the other half is its mirror image such that $F(w)$ diverges at $w \approx - 2.94$.

The Gaussian curvature of wave front with constant $u$ is given in Appendix A. In this particular case, it can be simply obtained from formula (B2) of Appendix B of Ref.~\cite{Bini:2018gbq}, namely, 
\beq \label{N23}
K_G = -\frac{1}{4}\,e^{-B} \left[2\,C_{xx} - (B_x - C_x) C_x\right]\,,
\eeq
or, since the gravitational potentials are all functions of $w$,  
\beq \label{N24}
K_G = -\frac{p^2}{4}\,e^{-B} \left[2\,C'' - (B' - C') C'\right]\,.
\eeq
Using Eq.~\eqref{N5}, we find
\beq \label{N25}
K_G = -\frac{p^2}{4}\,A'^2\,e^{-B} < 0\,.
\eeq
Therefore, in general, our TGWs have wave fronts with negative Gaussian curvature.  

In the following subsections, we will consider some special parameter values.

\subsection{$\Lambda = 0$}

It is a consequence of Eq.~\eqref{N3} that for $\Lambda = 0$ we have $A' = -2\,C'$, since $A'\ne 0$ by assumption. Moreover, $A = -2\,C$ plus a constant that can be absorbed in a redefinition of the $y$ coordinate.  It then follows from Eq.~\eqref{N8} that
\beq \label{N26}
e^B = \frac{1}{k_A^2}\,F\,F'^2\,.
\eeq
Furthermore, Eq.~\eqref{N10} implies  $k_C = 3\, k_A$ in this case. The wave front has negative Gaussian curvature given by $K_G =  -k_A^2\,p^2/F^3$, where $F$ is a solution of the differential equation
\beq \label{N27}
2\,F'' - \beta = \frac{F'^2}{F}\,.
\eeq
This equation and its first integral can be obtained from Eqs.~\eqref{N19} and~\eqref{N21} for $\alpha = 0$, respectively. Let us note that for $k_B = 0$, we have $\beta = 0$ and $\sqrt{F}$ depends linearly on $w$. In this special case, the spacetime metric takes the form
\beq \label{N28}
ds^2 = w^4 (-dt^2 + dz^2 + dx^2) + w^{-2} dy^2\,,
\eeq
which is of Petrov type D  and essentially coincides with the metric discussed in Refs.~\cite{Bini:2018iyu, Rosquist:2018ore}.

\subsection{$k_C = 0$}

If we assume that $k_C = 0$, the difference between $A$ and $C$ must be a constant that can be absorbed in the redefinition of the $y$ coordinate. Therefore, we set $A = C$. It then follows from Eq.~\eqref{N8} that
\beq \label{N29}
e^{\frac{1}{2}\,B} = \frac{1}{k_A}\,\frac{F'}{F}\,.
\eeq
Substituting this relation in Eq.~\eqref{N3}, we find
\beq \label{N30}
\Lambda = 3 K_G = - 3\,k_A^2\,p^2 < 0\,.
\eeq
The metric coefficients are determined in this case from the differential equation for $F$, namely,
\beq \label{N31}
2\,F'' - \beta = 4\,\frac{F'^2}{F}\,,
\eeq
which also follows from Eq.~\eqref{N19} by writing it as
\beq \label{N32}
2\,F'' - \beta = \frac{F'^2}{F}\,\frac{1+4\,\alpha\,F^3}{1+\alpha\,F^3}\,
\eeq
and formally letting $\alpha$ go to infinity. The first integral of Eq.~\eqref{N31} is given by 
\beq \label{N33}
F'^2 = \beta_0\,F^4  - \frac{1}{3}\,\beta\,F\,,
\eeq
where $\beta_0$ is an integration constant. 
Finally, let us mention that if $k_B = 0$ as well, then $\beta = 0$ and via constant rescalings of spacetime coordinates and parameters $(s, p)$, the spacetime metric can be rendered conformally flat; that is, 
\beq \label{N34}
ds^2 = w^{-2}\, \eta_{\mu \nu}\, dx^\mu dx^\nu\,, \qquad  w = s\,u \pm \sqrt{-\Lambda/3}\,x\,,
\eeq
where $(\eta_{\mu \nu}) =$diag$(-1, 1, 1, 1)$ is the Minkowski metric tensor.  A simple generalization of this conformally flat TGW is derived at the end of Section IV.

\section{Addition of a Cross Term}

Let us now continue the general approach adopted in Section II with the addition of a cross term and look for TGW solutions with metrics of the form
\begin{equation}\label{T1}
 ds^2 = - e^{A(w)}(dt^2-dz^2) + e^{B(w)}\,dx^2 + 2\,h(w) dx\,dy + e^{C(w)}\,dy^2\,,
\end{equation}
where the coordinate admissibility condition is in this case 
\begin{equation}\label{T2}
 f :=  e^{B + C} - h^2 > 0\,.
\end{equation}
We note that $w = s\,u + p\,x$; hence, the simple coordinate transformation $y \mapsto -y$ changes the overall sign of the cross term $h$, but otherwise leaves the metric invariant. 

For the explicit determination of the field equations, we introduce the standard null coordinates and write $dt^2-dz^2 = du\,dv$ in Eq.~\eqref{T1} and then work out the consequences of $R_{\mu \nu} = \Lambda \,g_{\mu \nu}$, cf. Appendix B. As before, $R_{vv} = R_{vx} = R_{vy} = 0$  are trivially satisfied by symmetry. We then have four inhomogeneous equations depending on the $\Lambda$ term, namely, $R_{xy} =  \Lambda\,h$, $R_{yy}  =  \Lambda\,\exp{(C)}$, $R_{uv}  = -(\Lambda/2)\,\exp{(A)}$ and $R_{xx}  =  \Lambda\,\exp{(B)}$, as well as three homogeneous equations 
$R_{ux} = R_{uy} = R_{uu} = 0$. The results of this section depend crucially on the assumption that
\begin{equation}\label{T2a}
 h(w) \ne 0\,.
\end{equation}

Let us start with the inhomogeneous field equations. Using Eq.~\eqref{T2} and its derivative, $R_{xy} =  \Lambda\,h$ and  $R_{yy}  =  \Lambda\,\exp{(C)}$ lead to the same equation which can be expressed as
\begin{equation}\label{T3}
 C'\,f' - 2\,f(C'' + C'^2 + C'\,A') = 4 \lambda f^2 e^{-C}\,,
\end{equation}
where $\lambda$ is the reduced cosmological constant defined in Eq.~\eqref{N15}. In the same way, from $R_{uv}  = -(\Lambda/2)\,\exp{(A)}$ we get
\begin{equation}\label{T4}
 A'\,f' - 2\,f(A'' + A'^2 + C'\,A') = 4 \lambda f^2 e^{-C}\,.
\end{equation}
Finally, $R_{xx}  =  \Lambda\,\exp{(B)}$ implies
\begin{equation}\label{T5}
 [C'\,f' - 2\,f(C'' + C'^2 + C'\,A')] e^C -2\,f^2 \left(2\,A'' + A'^2 -\frac{f'}{f} A'\right)e^{-B} = 4 \lambda f^2\,.
\end{equation}
If in this equation the part proportional to $\exp{(-B)}$ vanishes, we recover Eq.~\eqref{T3}. Therefore  Eqs.~\eqref{T3} and~\eqref{T5} imply 
\begin{equation}\label{T5a}
 2\,A'' + A'^2 -\frac{f'}{f} A' = 0\,,
\end{equation}
which can be simply integrated. The result is
\begin{equation}\label{T6}
 f = \ell_0\,A'^2 e^A\,,
\end{equation}
where $\ell_0 > 0$ is a constant of integration. Moreover, subtracting Eq.~\eqref{T3} from Eq.~\eqref{T4} results in
\begin{equation}\label{T7}
 f = \ell_1\,(A'-C')^2 e^{2(A+C)}\,,
\end{equation}
where $\ell_1 > 0$ is another constant of integration. From Eqs.~\eqref{T6} and~\eqref{T7}, we find
\begin{equation}\label{T8}
 \pm \sqrt{\frac{\ell_0}{\ell_1}}\, A' e^{-3A/2} = (A'-C') e^{-(A-C)}\,,
\end{equation}
which can be simply integrated to yield
\begin{equation}\label{T9}
e^C =  \pm \frac{2}{3}\,\sqrt{\frac{\ell_0}{\ell_1}}\, e^{-A/2} +\ell_2\,e^A\,,
\end{equation}
where $\ell_2$ is an integration constant. Employing Eq.~\eqref{T7} in Eq.~\eqref{T3}, we find
\begin{equation}\label{T10}
\left(\frac{C'}{A'}\right)' \,\left(\frac{C'}{A'}-1\right)^{-3} = 2\,\lambda \,\ell_1\,A' e^{2\,A + C}\,.
\end{equation}
Calculating $C'/A'$ via Eq.~\eqref{T9} and substituting the result in Eq.~\eqref{T10}, we get a formula for the reduced cosmological constant, namely, 
\begin{equation}\label{T11}
 \lambda = -\frac{3}{4} \,\frac{\ell_2}{\ell_0}\,.
\end{equation}

It remains to investigate the homogeneous equations. Let us start with $R_{uy} = 0$. As before, employing Eq.~\eqref{T2} and its derivative in $R_{uy} = 0$ leads to simplifications that turn this field equation into 
\begin{equation}\label{T12}
 2\,f\,[(C'' + C'^2) h - h''] - f'\,(C'\,h - h') = 0\,, 
\end{equation}
which can be simply integrated to yield
\begin{equation}\label{T13}
 f = \ell_3\,(C'\,h - h')^2 e^{2C}\,,
\end{equation}
where $\ell_3 > 0$ is a constant of integration. From Eqs.~\eqref{T7} and~\eqref{T13}, one can derive a differential equation whose solution is 
\begin{equation}\label{T14}
e^A =  \sqrt{\frac{\ell_3}{\ell_1}} \,\, h +\ell_4\,e^C\,,
\end{equation}
where we have written $h$ instead of $\pm\,h$ or $\mp \,h$, since the overall sign of $h$ can be changed via the coordinate transformation $y \mapsto -y$, and $\ell_4$ is an integration constant. Next, using Eq.~\eqref{T9} in Eq.~\eqref{T14}, we find
\begin{equation}\label{T15}
h =  L_0\, e^{-A/2} + L_1\,e^A\,,\quad L_0 := \mp \frac{2}{3}\,\ell_4\,\sqrt{\frac{\ell_0}{\ell_3}}\,, \quad L_1:= \sqrt{\frac{\ell_1}{\ell_3}}(1-\ell_2\,\ell_4)\,.
\end{equation}
Let us now consider $R_{ux} = 0$, which reduces in the same way to 
\begin{equation}\label{T16}
 (f + h^2)\,\left[2\,(C'' + C'^2)  -  C' \,\frac{f'}{f}\right] + hh'\,\frac{f'}{f}-2\,hh'' -A'f' + 2\,A''f = 0\,. 
\end{equation}
Substituting Eq.~\eqref{T12} in Eq.~\eqref{T16} and using Eq.~\eqref{T5a}, we find after some algebra
\begin{equation}\label{T17}
2\,(C'' + C'^2)  -  C' \,\frac{f'}{f} = A'^2\,. 
\end{equation}
This result is in fact a simple consequence of Eqs.~\eqref{T5a} and~\eqref{T9}. 

The last field equation to consider is then $R_{uu} = 0$. In the same manner as before, this field equation reduces to 
\begin{equation}\label{T18}
f'' -\frac{1}{2} \frac{f'^2}{f} - A'\,f' = B'C'e^{B+C} - h'^2\,. 
\end{equation}
From Eq.~\eqref{T2} and its derivative, we find
\begin{equation}\label{T19}
 B' = \frac{f' + 2\,hh'}{f+h^2} - C'\,, 
\end{equation}
so that the field equation under consideration takes the form
\begin{equation}\label{T20}
f'' -\frac{1}{2} \frac{f'^2}{f} - A'\,f' = C' (f' + 2\,hh') - C'^2 (f+h^2) - h'^2\,. 
\end{equation}
It is useful at this point to introduce the function $F(w)$, given as in Eq.~\eqref{N11} of Section II by $F = \exp{(A/2)}$. Then,
\begin{equation}\label{T21}
f = 4\, \ell_0 \,F'^2\,, \qquad A' = 2 \,\frac{F'}{F}\,. 
\end{equation}
Employing Eq.~\eqref{T9} for $C$ and Eq.~\eqref{T15} for $h$, field Eq.~\eqref{T20} reduces, after much algebra, to the autonomous differential equation
\begin{equation}\label{T22}
\frac{F'''}{F'} + W_1(F) \frac{F''}{F} + \frac{1}{2}\,W_2(F) \frac{F'^2}{F^2} + W_3(F) = 0\,, 
\end{equation}
where $W_{i}$, $i = 1,2,3$, are given by
\begin{equation}\label{T23}
W_1 = -  \frac{1 + 4\,\tilde{\lambda}\,F^3}{1 + \tilde{\lambda}\,F^3}\,, \quad W_2 = \left(\frac{1 - 2\,\tilde{\lambda}\,F^3}{1 + \tilde{\lambda}\,F^3}\right)^2\,, \quad W_3 =\frac{9\,\ell_1}{8\,\ell_0 \ell_3} \left(\frac{F}{1 + \tilde{\lambda}\,F^3}\right)^2\,. 
\end{equation}
Here, $\tilde{\lambda}$ is proportional to the cosmological constant and is given by
\begin{equation}\label{T24}
\tilde{\lambda} := \mp 2 \,\lambda \sqrt{\ell_0 \ell_1}\,. 
\end{equation}
Let us introduce $\Psi$ given by
\begin{equation}\label{T25}
\Psi:= \frac{1}{2}\,F'^2\,; 
\end{equation}
then, Eq.~\eqref{T22} can be written as 
\begin{equation}\label{T26}
F^2\,\frac{d^2 \Psi}{dF^2} + W_1(F) F\frac{d\Psi}{dF} + W_2(F) \Psi +  F^2\,W_3(F) = 0\,.
\end{equation}
In principle, from this linear inhomogeneous second-order ordinary differential equation we can determine $A(w)$ and hence the other metric functions for these TGWs with a cosmological constant. That is, given an appropriate solution of Eq.~\eqref{T26} for $A(w)$, Eqs.~\eqref{T2}, \eqref{T6}, \eqref{T9} and~\eqref{T15} can be used to find the corresponding spacetime metric. We note that the homogeneous part of Eq.~\eqref{T26} has regular singular points at $F = 0, \infty$ and $(-\tilde{\lambda})^{-1/3}$ for $\tilde{\lambda} < 0$.

These TGWs have Weyl curvature tensors that are algebraically special and of type II in the Petrov classification. The Kretschmann invariant is positive in this case and is given by
\beq \label{T26b}
\mathcal{K}_{III} = \frac{8}{3}\, \Lambda^2 + \frac{p^4}{3\,\ell_0\,\ell_1}\,\frac{1}{F^6(w)}\,,
\eeq
so that, as expected, we have a curvature singularity at $F = 0$. 

\begin{figure}
\includegraphics[scale=0.4]{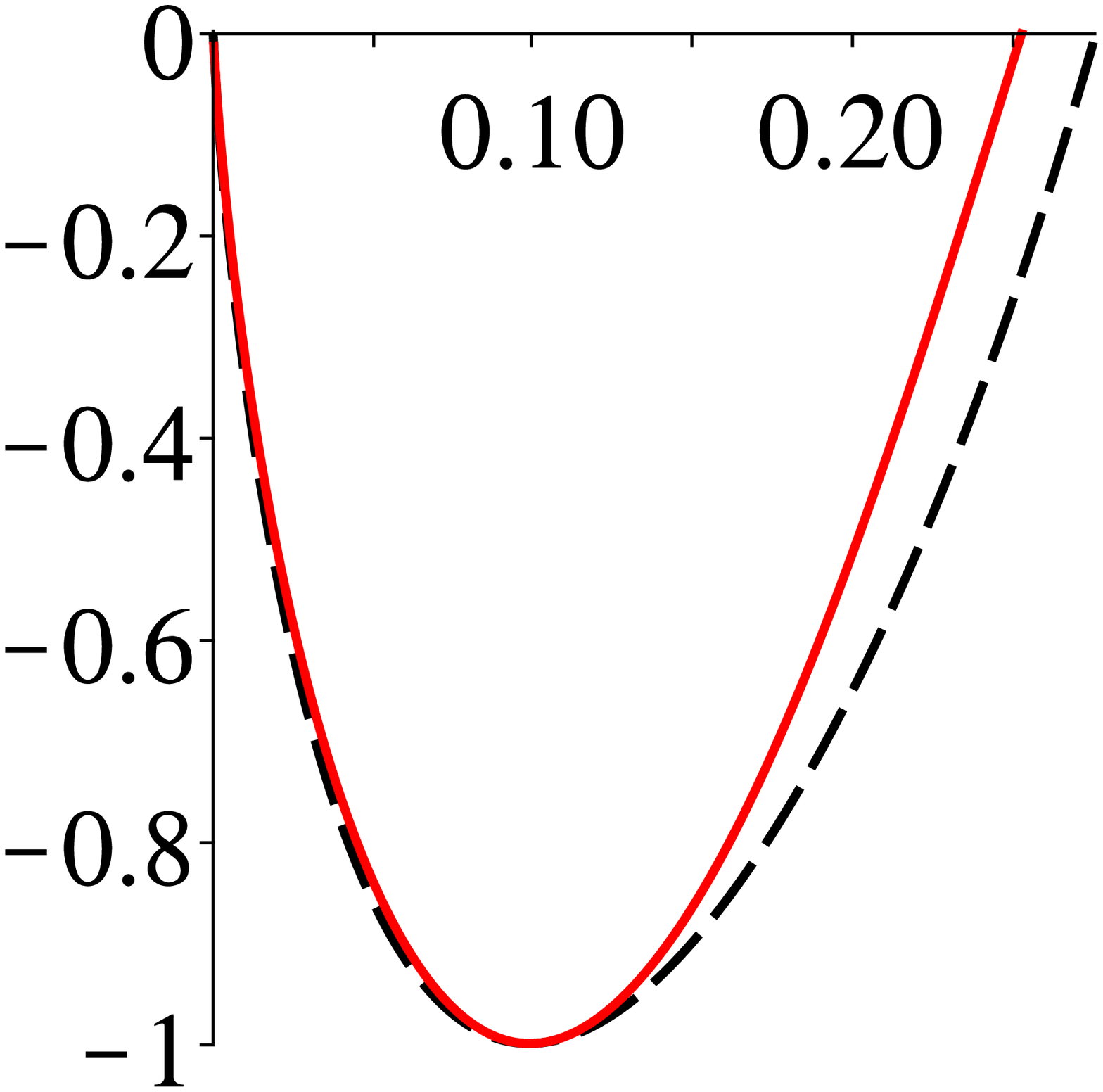} \hspace{0.3cm}
\includegraphics[scale=0.4]{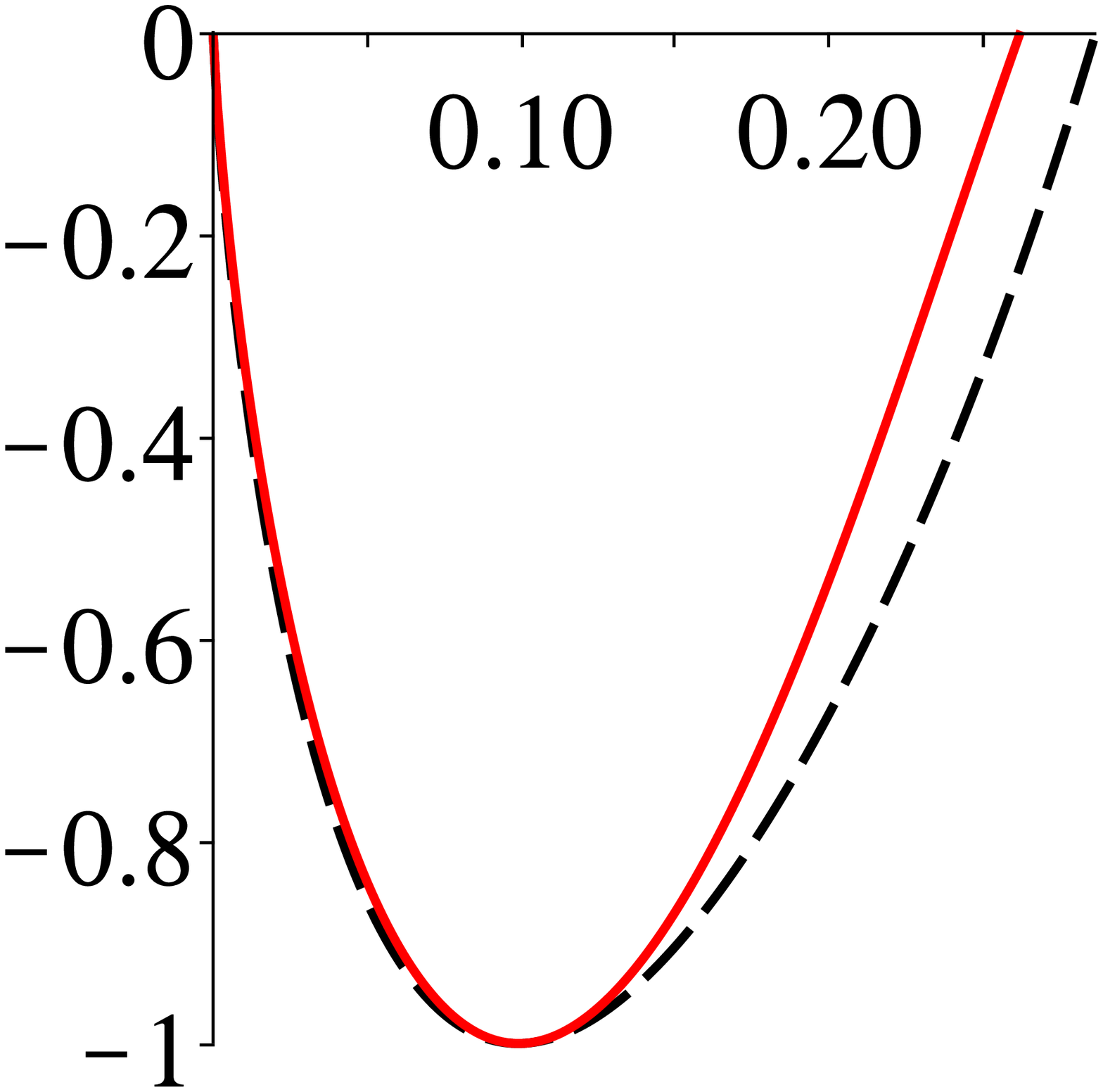}
\caption{\label{figIII} Left panel: Plot of the effective potential $\Upsilon$ versus $F$ for zero total energy and parameter values $9\ell_1/(8 \ell_0 \ell_3) = 100$ and $\tilde{\lambda} = \pm\, 20$. The solid red curve is for $\Lambda>0$ (i.e. $\tilde{\lambda} = -20$), while the dashed black curve is for $\Lambda <0$ (i.e. $\tilde{\lambda} = 20$).  Right panel: Plot of the effective potential $\Upsilon$ versus $F$ for zero total energy and  parameter values $9\ell_1/(8 \ell_0 \ell_3) = -100$ and $\tilde{\lambda} = \pm \,20$.  The solid red curve is for $\Lambda>0$ (i.e. $\tilde{\lambda} = -20$), while the dashed black curve is for $\Lambda <0$ (i.e. $\tilde{\lambda} = 20$). } 
\end{figure}

To investigate the general behavior of the spacetimes under consideration here, we note that, as before, we can think of the motion of a one-dimensional classical particle with zero total energy that has kinetic energy $\Psi:= \frac{1}{2}\,F'^2$ and potential energy $\Upsilon = -\Psi$. To find $\Upsilon$ versus $F$, we must investigate the nature of solutions of 
\begin{equation}\label{T26a}
F^2\,\frac{d^2 \Upsilon}{dF^2} + W_1(F) F\frac{d\Upsilon}{dF} + W_2(F) \Upsilon -  F^2\,W_3(F) = 0\,,
\end{equation}
which depend upon two parameters, namely, $9\ell_1/(8 \ell_0 \ell_3)$ and $\tilde{\lambda}$. For  $9\ell_1/(8 \ell_0 \ell_3) = \pm \,100$ and $\tilde{\lambda} = \pm \,20$, the results are presented in Figure~\ref{figIII}, where Eq.~\eqref{T26a} has been integrated with initial conditions that at $F = 0.1$,  $\Upsilon (0.1)= -1$ and $d\Upsilon/dF (0.1) = 0.1$. We have checked that the general shapes of the plots in Figure~\ref{figIII} are insensitive to the parameters of the system as well as the initial conditions of the integration. 
It follows from the effective potential method and the results of  Figure~\ref{figIII} that the motion is confined between  two turning points and the behavior of $F(w)$ as a function of $w$ is therefore periodic. 

The Gaussian curvature of the wave front for these TGWs can be calculated using the formula given in Appendix A. With $w = s\,u +p\,x$, where $u = u_0$ is a constant, we find
\begin{equation}\label{T27}
K_G = \frac{p^2\,e^C}{4\,f^2}\,[C'f' -2\,f(C''+C'^2)]\,, 
\end{equation}
which simplifies via Eq.~\eqref{T17} and the result is
\begin{equation}\label{T28}
K_G = -\frac{p^2\,A'^2\,e^C}{4\,f} < 0\,, 
\end{equation}
so that, as before, the Gaussian curvature of the wave front is negative. 

Returning to Eq.~\eqref{T26}, let us note that this equation simplifies considerably in the absence of a cosmological constant (i.e., $\tilde{\lambda} = 0$) . In fact, the general solution of this equation can be expressed as
\begin{equation}\label{T29}
\Psi  = (\mu_{1}  +  \mu_{2} \ln F)\,F - \frac{\ell_1}{8\,\ell_0 \ell_3} F^4\,, 
\end{equation}
where $\mu_1$ and $\mu_2$ are integration constants and $F'^2 = 2\,\Psi$.

\section{A Simple Generalization}

We consider a metric of the form
\begin{equation}\label{S1}
 ds^2 = - e^{A(w)}\,du\,dv + e^{B(w)}\,dx^2 + e^{C(w)}\,dy^2\,,
\end{equation}
where $x$ and $y$ are now treated on the same footing, namely, 
\begin{equation}\label{S2}
 w = s\,u + p\,x + q\,y\,, \qquad s \ne 0\,, \qquad p \ne 0 \,, \qquad q \ne 0\,.
\end{equation}
This spacetime contains three Killing vector fields as well; that is, $\partial_t + \partial_z$,  $p\, \partial_t - s\, \partial_x$ and  $q\, \partial_t - s\,\partial_y$.
The field equations presented in Appendix B contain three trivial ones that simply vanish by symmetry, namely, $R_{vv} = R_{vx} = R_{vy} = 0$. The others include four homogeneous  and three inhomogeneous field equations, the latter involving the cosmological constant. The four homogeneous equations, namely, $R_{xy} = 0$, $R_{uy} = 0$, $R_{ux} = 0$ and $R_{uu} = 0$ can be expressed as 
\beq \label{S3}
2\,A'' +A'^2 = A'(B'+C')\,,
\eeq
\beq \label{S4}
2\,B'' +B'^2 + 2\,A'' = A'(B'+C') + B'\,C'\,,
\eeq
\beq \label{S5}
2\,C'' + C'^2 + 2\,A'' = A'(B'+C') + B'\,C'\,
\eeq
and
\beq \label{S6}
2\,B''+B'^2 + 2\,C'' + C'^2 = 2\,A' (B'+C')\,,
\eeq
respectively. Inspection of Eqs.~\eqref{S4} and~\eqref{S5} reveals the symmetry between $B$ and $C$, so that $2\,B''+B'^2 = 2\,C'' + C'^2$; then the other equations imply  
\beq \label{S7}
2\,A'' +A'^2 = 2\,B''+B'^2 = 2\,C'' + C'^2 = A' (B'+C')\,,\qquad  2\,A'' = B'\,C'\,.
\eeq
Substituting $2\,A'' = B'\,C'$ in Eq.~\eqref{S3}, we find
\beq \label{S8}
(A' - B')(A' - C') = 0\,.
\eeq
Thus either $A' = B'$ or $A' = C'$; however, the symmetry between $B$ and $C$ implies that it is sufficient to consider one of these; therefore, we choose the case $A' = B'$. Moreover, we can henceforth simply set
\beq \label{S9}
A = B\,,
\eeq
since the constant of integration can always be absorbed in the redefinition of the advanced null coordinate $v$. With $A = B$, Eq.~\eqref{S3} can now be integrated and we find
\beq \label{S10}
e^C = \hat{k}_C^2\, A'^2\,,
\eeq
where $\hat{k}_C^2$ is a nonzero integration constant. By a simple rescaling of the $y$ coordinate and parameter $q$, namely, $(y\,\hat{k}_C, q/\hat{k}_C) \to (y, q)$, it is possible to set $\hat{k}_C^2 = 1$ with no loss in generality. Using Eq.~\eqref{S10} in $2\,C'' + C'^2 = 2\,A'' +A'^2$ results in an ordinary differential  equation for $A(w)$,
\beq \label{S11}
4A''' - 2 A'\,A'' - A'^3 = 0\,.
\eeq

Next, the three inhomogeneous equations, namely, $R_{uv} = - (\Lambda/2)\,\exp{(A)}$,  $R_{xx} = \Lambda\,\exp{(B)}$ and $R_{yy} = \Lambda\,\exp{(C)}$, all reduce to the same equation when we employ Eqs.~\eqref{S7} and~\eqref{S9}, namely, 
\beq \label{S12}
(4A'' + A'^2)\,e^{-A} = 3\,\Sigma_0\,, \qquad \Sigma_0 = - \frac{4\,\Lambda + 3\,q^2}{3\,p^2}\,.
\eeq
Remarkably, Eq.~\eqref{S12} turns out to be a first integral of Eq.~\eqref{S11} and $\Sigma_0$ is simply an integration constant. It is possible to integrate Eq.~\eqref{S12} once and the result is the autonomous differential equation
\beq \label{S13}
A'^2 = \Sigma_0\,e^A + \Pi_0\,e^{-\frac{1}{2} A}\,, 
\eeq
where $\Pi_0$ is another integration constant. Equation~\eqref{S13} can be solved by quadrature; that is,
\beq \label{S14}
\int^{e^{\frac{1}{4} A}} \frac{d\zeta}{\sqrt{\Pi_0 +\Sigma_0\,\zeta^6}} = \pm \frac{1}{4} \,w\,. 
\eeq

To investigate the character of this class of TGW solutions, it is useful to define $F:= \exp{(A/2)}$ as in Section II and write Eq.~\eqref{S13} as 
\beq \label{S14a}
4\,F'^2  + \mathcal{V}(F) = 0\,, \qquad  \mathcal{V}(F) = - (\Sigma_0\,F^4 + \Pi_0\,F) \le 0\,.
\eeq
In terms of $F > 0$, the metric functions are
\beq \label{S14b}
e^A = e^B = F^2\,, \qquad e^C = \Sigma_0 \,F^2 + \frac{\Pi_0}{F}\,.
\eeq
These general solutions have Weyl curvature tensors that are algebraically special and of type D in the Petrov classification.The Kretschmann invariant for this class is positive and is  given by
\beq \label{S14c}
\mathcal{K}_{IV} = \frac{8}{3}\, \Lambda^2 + \frac{3}{4}\,\frac{\Pi_0^2\,p^4}{F^6(w)}\,,
\eeq
so that, as in Sections II and III, the curvature singularity occurs at $F = 0$. Let us note that the curvature singularity disappears for $\Pi_0 = 0$. In this case, we must have $\Sigma_0 > 0$ and by constant rescalings of the $y$ coordinate as well as parameters $(s, p, q)$ the solution reduces to the special conformally flat (AdS) spacetime discussed in detail in the subsection below and Section V.  Henceforth, we assume $\Pi_0 \ne 0$. If $\Sigma_0 = 0$, then $\Lambda = - 3\,q^2/4 < 0$ and we obtain, after constant rescalings of the spacetime coordinates as well as parameter $q$, a natural generalization of solution~\eqref{N28} for a negative cosmological constant. 

\begin{figure}
\includegraphics[scale=0.5]{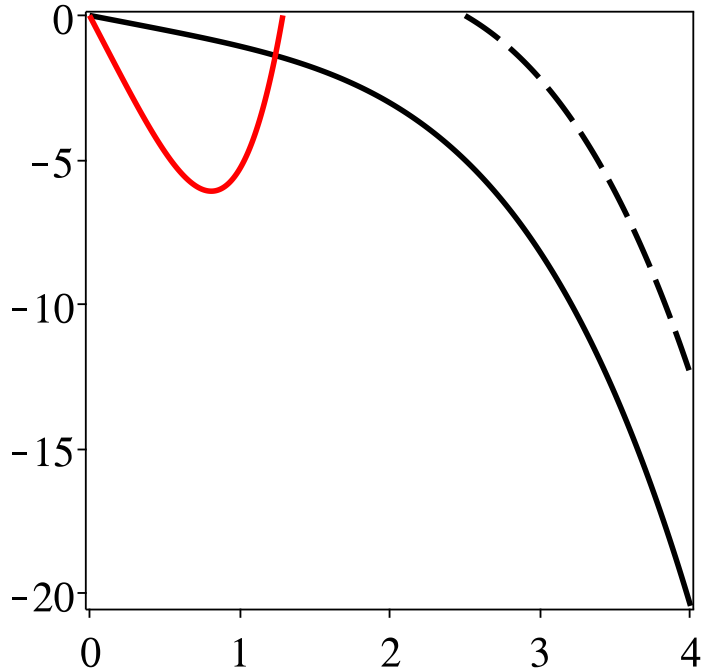}
\caption{\label{figIV} Plot of the effective potential $\mathcal{V}$ versus $F \ge 0$ for zero total energy.  We note that $\mathcal{V}(F)$ given in Eq.~\eqref{S14a}  vanishes at $F = 0$ and $F = (-\Pi_0/\Sigma_0)^{1/3}$. We choose $p = 1$ and $q = 0.45$. The red curve, which represents $10\,\mathcal{V}$ versus $F$ to aid visualization, is for $\Pi_0 = 1$ and $\Sigma_0 = - 0.47$, so that $\Lambda \approx 0.2$. The solid black curve is for $\Pi_0 = 1$ and $\Sigma_0 = 0.07$, so that $\Lambda \approx - 0.2$. Finally, the dashed black curve is for $\Pi_0 = -1$ and $\Sigma_0 = 0.07$, so that $\Lambda \approx - 0.2$.}
 \end{figure}

As before, we interpret Eq.~\eqref{S14a} in terms of one-dimensional motion of a classical particle of net energy zero that has kinetic energy $4\,(dF/dw)^2$ and potential energy $\mathcal{V}(F)$ illustrated in Figure~\ref{figIV}. For $\Sigma_0 > 0$, $\mathcal{V}(F)$ diverges as $F \to \infty$. The behavior of $F(w)$ versus $w$ is essentially the same as described in Section II, cf. the right panel of Figure~\ref{figII}. 

Finally, the wave front has negative Gaussian curvature. Using the result given in Appendix A or formula (B2) of Appendix B of Ref.~\cite{Bini:2018gbq}, we find for the Gaussian curvature  of the wave front ($u=$ constant),
\beq \label{S15}
K_G = -\frac{1}{4}\,A'^2\, \left( p^2 \,e^{-B} + q^2\,e^{-C}\right) < 0\,. 
\eeq

In connection with the possibility of the addition of a cross term in this case, we mention that the analytic treatment of the problem appears to be prohibitively complicated. This conclusion is based on a close inspection of the field equations given in Appendix B.

\subsection{Conformally Flat Solution}

Let us assume that $C = A$ in Eq.~\eqref{S10}, so that the spacetime metric is conformally flat. This means via Eqs.~\eqref{S12} and~\eqref{S13}  that $\Sigma_0 = 1$, $\Pi_0 = 0$ and 
\beq \label{S16}
p^2 + q^2  = - \frac{4}{3}\,\Lambda\,,
\eeq
which is possible if the cosmological constant is \emph{negative}. We can recast this solution into the form $ds^2 = \Omega^2\, \eta_{\mu \nu}\,dx^\mu\,dx^\nu$, where $\Omega^{-1} = \hat{s}\,u + \hat{p}\,x + \hat{q}\,y$ such that
\beq \label{S17}
(\hat{s}, \hat{p}, \hat{q}) = \frac{1}{2} (s, p, q)\,.
\eeq
For $q = 0$, this solution reduces to the simple conformally flat solution~\eqref{N34} we found in Section II. It is convenient to introduce an angle $\theta$, $0 \le \theta < 2\,\pi$, such that
\beq \label{S18}
\hat{p} = (-\Lambda/3)^{1/2}\,\cos\theta\,, \qquad \hat{q} = (-\Lambda/3)^{1/2}\,\sin\theta\,.
\eeq
We show  in Appendix C that the TGW spacetime under discussion here is indeed the \emph{unique} solution of $R_{\mu \nu } = \Lambda\,g_{\mu \nu}$ that is conformally flat and represents a unidirectional gravitational wave. Henceforward, we drop the hats on $(s, p, q)$ for the sake of simplicity. 

Conformally flat plane wave spacetimes have been the subject of extensive investigations, see, for instance, Ref.~\cite{R1}, p. 603. The corresponding energy-momentum tensor is of the null fluid type, which can be interpreted either in terms of null dust or a pure (null) electromagnetic radiation field~\cite{Skea, Griffiths:1998sp}. An example of the latter situation has been discussed in detail, in connection with the phenomenon of cosmic jets~\cite{Chicone:2010xr},  in Section IV of Ref.~\cite{Bini:2014esa}. Conformally flat Kundt solutions with a cosmological constant are treated in Ref.~\cite{Griffiths:2009dfa}, Section 18.3.3, p. 344.

\section{Conformally Flat TGW due to a Negative $\Lambda$}

The purpose of this section is to investigate in more detail the Petrov type O solution that we found in the previous section, namely, 
\beq \label{L1}
g_{\mu \nu}  =  \Omega^2 \,\eta_{\mu \nu}\,, \qquad  \Omega^{-1} = s\,u  + \varpi\,\cos\theta\,x + \varpi\,\sin\theta\,y\,,
\eeq
where  $u = t-z$ is the retarded null coordinate, $s \ne 0$ and $\theta$ are constant parameters  and $\varpi > 0$ is given by
\beq \label{L2}
\varpi :=  \left(-\frac{\Lambda}{3} \right)^{1/2}\,.
\eeq

For a conformally flat spacetime, the Weyl curvature tensor $C_{\mu \nu \rho \sigma}$ vanishes. Therefore, a Ricci-flat spacetime representing a nonlinear gravitational wave cannot be conformally flat; otherwise, the Riemann curvature tensor would completely vanish. Thus the existence of our solution is purely due to the presence of the cosmological constant $\Lambda < 0$. Indeed, for $\Lambda = 0$,  our metric in $(u, v, x, y)$ coordinates under  rescalings of the spacetime coordinates via $(u, v, x, y) \mapsto (u^{-1}, -s^2\,v, s\,x, s\,y)$  reduces to  
\beq \label{L3}
ds^2 = - du\,dv +u^2\, (dx^2+dy^2)\,, 
\eeq
which is simply flat as demonstrated at the end of Appendix B of Ref.~\cite{Bini:2017qnd}.

The curvature tensor for the spacetime under consideration reduces to 
\beq \label{L4}
R_{\mu \nu \rho \sigma} = \frac{1}{3} \Lambda (g_{\mu \rho}\,g_{\nu \sigma} - g_{\mu \sigma}\,g_{\nu \rho})\,,
\eeq
which means that the spacetime has constant negative curvature and is thus part of the anti-de Sitter manifold~\cite{R1, Griffiths:2009dfa, HE}.  Furthermore, the wave front in this case has constant negative Gaussian curvature 
\beq \label{L5}
K_G = \frac{1}{3}\, \Lambda < 0\,,
\eeq 
cf. Appendix A.  It is interesting to examine the connection of TGW solution~\eqref{L1} with the anti-de Sitter solution. This is done in the rest of this section.

\subsection{Accelerated System in AdS Spacetime}

Imagine a congruence of accelerated observers in anti-de Sitter spacetime. In coordinates $x^\mu = (t, x, y, z)$ adapted to these observers such that they are spatially at rest  in these coordinates, the uniformly curved anti-de Sitter spacetime appears as a conformally flat TGW with metric $ds^2 = \Omega^2 \eta_{\mu \nu} \,dx^\mu dx^\nu$, where $1/\Omega = b \cdot x = \eta_{\alpha \beta}\, b^\alpha x^\beta$ with a 4-vector $b^\alpha$,
\begin{equation}\label{U1}
 b^\alpha = (-s, p, q, -s)\,.
\end{equation}
In this subsection, the inner (dot) product is defined only via the Minkowski metric tensor; for instance, 
\begin{equation}\label{U2}
 b^2 = b \cdot b = p^2 + q^2 = -\Lambda/3 = \varpi^2\,.
\end{equation}

One can start from the standard form of the metric of anti-de Sitter spacetime and obtain the conformally flat TGW via coordinate transformations. We find it convenient to start instead with the TGW metric and transform it back to the standard AdS form. In this process, the first step involves the acceleration transformation $x \mapsto y$ given by
\begin{equation}\label{U3}
 x^\mu = \frac{y^\mu - a^\mu\,y^2}{1-2\,a\cdot y + a^2\,y^2}\,,
\end{equation}
where $a^\mu$ is a constant 4-vector. Under this transformation,
\begin{equation}\label{U4}
  \eta_{\mu \nu} \,dx^\mu dx^\nu = (1-2\,a\cdot y + a^2\,y^2)^{-2} \,  \eta_{\alpha \beta}\, dy^\alpha dy^\beta\,
\end{equation}
and
\begin{equation}\label{U5}
\Omega^2 = (1-2\,a\cdot y + a^2\,y^2)^{2} \, [b\cdot y- (a\cdot b)\, y^2]^{-2}\,,
\end{equation}
so that the TGW metric now has a different conformally flat form given by
\begin{equation}\label{U6}
ds^2 =  [b\cdot y- (a\cdot b)\, y^2]^{-2}\,\eta_{\alpha \beta} \,dy^\alpha dy^\beta\,.
\end{equation}

Next, under the spacetime translation $y \mapsto z$ via a constant 4-vector $\chi^\mu$, 
\begin{equation}\label{U7}
 y^\mu  = z^\mu + \chi^\mu\,,
\end{equation}
we find
\begin{equation}\label{U8}
ds^2 = \Phi^{-2}\,\eta_{\alpha \beta} \,dz^\alpha dz^\beta\,,
\end{equation}
where
\begin{equation}\label{U9}
\Phi =  b\cdot \chi- (a\cdot b)\, \chi^2 + \eta_{\alpha \beta} \,[b^\alpha - 2\,(a\cdot b)\, \chi^\alpha]\,z^\beta - (a\cdot b)\, z^2\,.
\end{equation}
We choose $\chi^\mu$ and $a^\mu$ such that
\begin{equation}\label{U10}
b^\alpha = -\frac{\Lambda}{6}\, \chi^\alpha\,, \qquad a\cdot b = -\frac{\Lambda}{12}\,.
\end{equation}
For instance, a possible choice for $a^\mu$ is $a^\mu = b^\mu /4$. Fixing  $a^\mu$ and $\chi^\mu$ in this way, we find 
\begin{equation}\label{U11}
\Phi =  1+ \frac{\Lambda}{12}\,z^2\,,
\end{equation}
which together with Eq.~\eqref{U8} constitutes a known conformally flat representation of the anti-de Sitter spacetime, cf. Eq.~(8.33) of Ref.~\cite{R1}.

Let us briefly digress here and mention that in Minkowski spacetime the group of acceleration transformations~\eqref{U3} is an Abelian subgroup of the spacetime conformal group, namely, the 15-parameter invariance group of the light cone. For instance, it follows from Eq.~\eqref{U3} that
\begin{equation}\label{U12}
x^2\,(1-2\,a\cdot y + a^2\,y^2) = y^2\,, \qquad (1-2\,a\cdot y + a^2\,y^2)^{-1} = 1+2\,a\cdot x + a^2\,x^2\,.
\end{equation}

The final step involves transforming Eqs.~\eqref{U8} and~\eqref{U11} to the standard AdS form. To simplify matters, let us express all lengths in these equations in units of $1/\varpi= (-\Lambda/3)^{-1/2}$. With this proviso as well as the introduction of spherical polar coordinates,
\begin{equation}\label{U13}
z^0 = \Theta\,, \qquad z^1 = R \sin\vartheta\,\cos\varphi\,, \qquad z^2 = R \sin\vartheta\,\sin\varphi\,, \qquad z^3 = R \cos\vartheta\,,
\end{equation}
we have
\begin{equation}\label{U14}
ds^2 = \frac{-d\Theta^2 + dR^2 + R^2\,(d\vartheta^2 + \sin^2\vartheta\,d\varphi^2)}{\left[1 + \frac{1}{4}\,(\Theta^2 - R^2)\right]^2}\,.
\end{equation}
The transformation of this metric to the standard AdS metric,
\begin{equation}\label{U15}
ds^2 = -( 1 + r^2) d{\hat t}^2 + \frac{dr^2}{1+r^2} + r^2\,(d\vartheta^2 + \sin^2\vartheta\,d\varphi^2)\,,
\end{equation}
can be simply accomplished via
\begin{equation}\label{U16}
\Theta = \tan({\hat u}/2)  +  \tan({\hat v}/2)\,, \qquad R =  - \tan({\hat u}/2)  +  \tan({\hat v}/2)\,,
\end{equation}
where
\begin{equation}\label{U17}
{\hat u} = {\hat t} - \arctan r\,, \qquad  {\hat v}   =  {\hat t} + \arctan r\,.
\end{equation}
More explicitly, we can write
\begin{equation}\label{U18}
\Theta = 2\,\frac{\sqrt{1+r^2}\,\sin{\hat t}}{1+ \sqrt{1+r^2}\,\cos{\hat t}}\,,\qquad R = 2\, \frac{r}{1+ \sqrt{1+r^2}\,\cos{\hat t}}\,.
\end{equation}

\section{Properties of the Conformally Flat TGW Solution}

It is interesting to study some of the main physical characteristics of the special conformally flat TGW solution~\eqref{L1} and compare them with previous results regarding TGWs~\cite{Bini:2018gbq, Bini:2018iyu, Rosquist:2018ore}.

\subsection{Timelike Geodesics}

Let us first investigate the motion of free test particles in this gravitational field. Null geodesics are conformally invariant; hence, null geodesics in this conformally flat TGW spacetime are the same as those in Minkowski spacetime. We therefore concentrate on timelike geodesics. There are three Killing vector fields in this spacetime, namely,
\beq \label{L6}
\partial_t + \partial_z\,, \qquad  \varpi \,\cos\theta\, \partial_t - s\, \partial_x\,, \qquad \varpi \,\sin\theta\, \partial_t - s\,\partial_y\,.
\eeq 
Thus there are three constants of timelike geodesic motion that can be obtained from projecting the 4-velocity vector of a free massive test particle, $\dot{x}^\mu = dx^\mu/d\eta$, on the Killing vector fields. Here $\eta$ is the proper time along the timelike  geodesic world line. We have 
\begin{equation}\label{L7}
\Omega^2 (\dot{t}-\dot{z}) = C_v\,, \qquad \Omega^2 (\varpi\,\cos\theta\,\dot{t} + s\, \dot{x}) = C_1\,, \qquad \Omega^2 (\varpi\,\sin\theta\,\dot{t} + s\, \dot{y}) = C_2\,, 
\end{equation}
where $C_v$, $C_1$ and $C_2$ are constants of the motion. Furthermore, the 4-velocity is a timelike unit vector; hence, $\Omega^2\,\eta_{\mu\nu}\,\dot{x}^\mu\, \dot{x}^\nu = - 1$.

It is convenient to take advantage of the circumstance that the geodesic equations of motion can be simply obtained from a Lagrangian of the form $(ds/d\eta)^2$. Therefore, we find
\begin{equation}\label{L8}
\frac{d}{d\eta}(\Omega^2\, \dot{t})= -s\,\Omega\,, \qquad \frac{d}{d\eta}(\Omega^2\, \dot{x})= \varpi\,\cos\theta\,\Omega\,, \qquad \frac{d}{d\eta}(\Omega^2\, \dot{y})= \varpi\,\sin\theta\,\Omega\,,
\end{equation}
etc. Let us write
\begin{equation}\label{L9}
\frac{d}{d\eta}(\Omega^{-1}) = \frac{d}{d\eta}(s\,u + \varpi\,\cos\theta\,x + \varpi\,\sin\theta \,y) =  s\,\frac{C_v}{\Omega^2} + \varpi\,\cos\theta\,\dot{x} + \varpi\,\sin\theta\,\dot{y}\,,
\end{equation}
where Eq.~\eqref{L7} has been used. Next, multiplying both sides of this equation with $\Omega^2 > 0$ and employing Eq.~\eqref{L8}, we find
\begin{equation}\label{L10}
\frac{d^2 \Omega}{d\eta^2} + \varpi^2 \,\Omega = 0\,,
\end{equation}
which has the general solution
\begin{equation}\label{L11}
 \Omega (\eta) = \Omega_0\,\cos[\varpi\,(\eta-\eta_0)]\,,
\end{equation}
where $\eta_0$ and $\Omega_0 = \Omega(\eta_0)$ are constants of integration. It is clear that as proper time $\eta$ increases monotonically from $\eta_0$, $\Omega(\eta)$ decreases monotonically and eventually approaches the singular value of zero at $\eta = \eta_0 +\pi / (2\,\varpi)$. This is a coordinate singularity and comes about due to the special spacetime coordinates adapted to these geodesic observers.  

It is now straightforward to use our result for $\Omega(\eta)$ in Eqs.~\eqref{L7} and~\eqref{L9} to find $\dot{x}^\mu(\eta)$. Integrating these results, we determine $x^\mu (\eta)$ for a timelike geodesic, which may be expressed as
\begin{equation}\label{L12}
x^\mu (\eta) - x^\mu (\eta_0) = \frac{\mathbb{C}_\mu}{\varpi^2\,\Omega_0^{2}}\, \tan[\varpi\,(\eta-\eta_0)] -\frac{\mathbb{D}_\mu}{\varpi^2\,\Omega_0}\,\left(1-\frac{1}{\cos[\varpi\,(\eta-\eta_0)]}\right)\,,
\end{equation}
where
\begin{equation}\label{L13}
\mathbb{C}_0 = \frac{s^2}{\varpi}\,C_v + C_1\,\cos\theta + C_2\,\sin\theta\,, \qquad \mathbb{C}_3 = \mathbb{C}_0 - \varpi\,C_v\,,
\end{equation}
\begin{equation}\label{L14}
\mathbb{C}_1 = -s\,C_v\,\cos\theta +\frac{\varpi}{s}\,(C_1\,\sin\theta - C_2\,\cos\theta)\,\sin\theta\,,
\end{equation}
\begin{equation} \label{L15}
\mathbb{C}_2 = -s\,C_v\,\sin\theta -\frac{\varpi}{s}\,(C_1\,\sin\theta - C_2\,\cos\theta)\,\cos\theta\,
\end{equation}
and 
\begin{equation}\label{L16}
 \mathbb{D}_0 = \mathbb{D}_3 = - s\,, \qquad \mathbb{D}_1 = \varpi\,\cos\theta\,, \qquad \mathbb{D}_2 = \varpi\,\sin\theta\,.
\end{equation}
The integration constants in these equations are related via
\begin{equation}\label{L17}
\frac{1}{\Omega_0} = s\,u(\eta_0) + \varpi \,[x(\eta_0)\,\cos\theta + y(\eta_0)\,\sin\theta]\,,
\end{equation}
where $u(\eta_0) = t(\eta_0) - z(\eta_0)$,  and 
\begin{equation}\label{L18}
\varpi^2\,\Omega_0^2 = (s^2-\varpi^2)\,C_v^2 + 2\,\varpi\,C_v\,( C_1\,\cos\theta + C_2\,\sin\theta) - \frac{\varpi^2}{s^2}\, (C_1\,\sin\theta - C_2\,\cos\theta)^2\,,
\end{equation}
which follows from $\Omega^2\,\eta_{\mu\nu}\,\dot{x}^\mu\, \dot{x}^\nu = - 1$.

\subsubsection{Cosmic Jet}

In certain dynamic spacetime regions, geodesics tend to line up, as measured by static fiducial observers, and thus produce a cosmic jet whose speed asymptotically approaches the speed of light~\cite{Chicone:2010xr, Bini:2014esa}. For \emph{plane} gravitational wave spacetimes, this cosmic jet property was first demonstrated  in Ref.~\cite{Bini:2014esa} and further studied in Ref.~\cite{Bini:2017qnd}. A plane gravitational wave admits parallel null rays, so that  the four principal null directions of the Weyl tensor coincide and are all parallel to the direction of propagation of the plane wave and hence perpendicular to the uniform wave front. With respect to the static observers in these spacetimes, timelike geodesics exhibit the cosmic jet property, where the jet motion is parallel to the direction of motion of the plane wave~\cite{Bini:2014esa, Bini:2017qnd}. However, the nonuniformity of the wave front in the case of nonplanar TGWs implies that the resulting cosmic jet direction is \emph{oblique} with respect to the direction of wave propagation~\cite{Bini:2017qnd}.  
It is interesting to investigate this property for the case under consideration here. To this end, imagine a congruence of timelike geodesics in our conformally flat TGW spacetime. We are interested in the motion of a member of this congruence at time $\eta$ with respect to a static observer spatially at rest in this spacetime. The natural tetrad frame of these static fiducial observers is given by
\begin{equation}\label{L19}
e^{\alpha}{}_{\hat \mu} = \frac{1}{\Omega}\,\delta^{\alpha}_{\mu}\,,
\end{equation}
where in $1/\Omega = s\,u +  \varpi\,\cos\theta\,x + \varpi\,\sin\theta\,y$, the spatial coordinates $x$, $y$ and $z$ are constants. These fiducial observers exist so long as $\Omega \ne 0$. Projecting $\dot{x}^\alpha = (\dot t, \dot x, \dot y, \dot z)$ upon $e_{\alpha}{}^{\hat \mu}$ at $x^\alpha (\eta)$  results in the instantaneous relation
\begin{equation}\label{L20}
\dot{x}^{\alpha}\,e_{\alpha}{}^{\hat \mu} = \Omega\,(\dot t, \dot x, \dot y, \dot z) = U^{\hat \mu} := \Gamma (1, V_x, V_y, V_z)\,,
\end{equation}
where $U^{\hat \mu}$ is the 4-velocity of the timelike geodesic as measured by the fiducial static observer. We find
\begin{equation}\label{L21}
U^{\hat \mu} = \frac{1}{\varpi\,\Omega_0}\,\frac{\mathbb{C}_\mu + \Omega_0\,\mathbb{D}_\mu\,\sin[\varpi\,(\eta-\eta_0)]}{\cos[\varpi\,(\eta-\eta_0)]}\,.
\end{equation}
It follows that as $\eta \to \eta_0 +\pi / (2\,\varpi)$, $\Gamma \to \infty$ and the oblique \emph{cosmic jet} is characterized by
\begin{equation}\label{L22}
V_x \to  \frac{\mathbb{C}_1 + \Omega_0\,\mathbb{D}_1}{\mathbb{C}_0 + \Omega_0\,\mathbb{D}_0}\,,\quad V_y \to  \frac{\mathbb{C}_2 + \Omega_0\,\mathbb{D}_2}{\mathbb{C}_0 + \Omega_0\,\mathbb{D}_0}\,,\quad V_z \to  \frac{\mathbb{C}_3 + \Omega_0\,\mathbb{D}_3}{\mathbb{C}_0 + \Omega_0\,\mathbb{D}_0}\,.
\end{equation} 
One can check using Eq.~\eqref{L18}  that indeed $V_x^2 + V_y^2 + V_z^2 \to 1$ as the cosmic jet develops.  
 
Tidal effects of our conformally flat TGWs are studied in the next subsection.

\subsection{Jacobi Equation}

Imagine a static observer that is at rest in space in our conformally flat TGW spacetime. The observer carries an orthonormal tetrad frame $e^{\mu}{}_{\hat \alpha}$ along its world line and uses this frame to set up a geodesic (Fermi) normal coordinate system in its neighborhood. The Fermi system is discussed in Appendix D.  We are interested in the motion of nearby geodesics with respect to the accelerated static observer that permanently occupies the origin of the Fermi coordinate system. This analysis is carried out in several steps. 

\subsubsection{Tetrad of the Static Observer}

The static fiducial observer has a natural tetrad frame $e^{\mu}{}_{\hat \alpha} = \Omega^{-1} \,\delta^\mu_\alpha$, where $\Omega \ne 0$. The world line of such an observer is given by
  $\bar{x}^\mu = (t, x_0, y_0, z_0)$, where
\begin{equation}\label{L23}
 \tau = \frac{1}{s}\,[\ln( s\,t + \varpi\,\cos\theta\,x_0 + \varpi\,\sin\theta\,y_0 - s\,z_0) -\ln( \varpi\,\cos\theta\,x_0 + \varpi\,\sin\theta\,y_0 - s\,z_0)]\,
\end{equation} 
 is the proper time of the static observer and we have assumed that $\tau = 0$ at $t=0$. Such observers are not geodesic; in fact, they are accelerated with  
\begin{equation}\label{L24}
\mathcal{A}^\mu = \frac{De^{\mu}{}_{\hat 0}}{d\tau} = - \varpi\,\cos\theta\,\, e^{\mu}{}_{\hat 1} - \varpi\,\sin\theta\,\, e^{\mu}{}_{\hat 2}+ s\,e^{\mu}{}_{\hat 3}\,. 
\end{equation} 
The static observer carries the spatial frame of the tetrad along its world line  for measurement purposes. It is straightforward to check that the spatial frame $e^{\mu}{}_{\hat i}$, for $i = 1, 2, 3$,  is indeed Fermi-Walker transported; that is, its components satisfy the equation of Fermi-Walker transport,  
\begin{equation}\label{L25}
\frac{d\mathbb{S}^\mu}{d\tau}+\Gamma^{\mu}_{\alpha \beta}\, e^{\alpha}{}_{\hat 0}\,\mathbb{S}^{\beta}= (\mathcal{A}\cdot \mathbb{S})\,e^{\mu}{}_{\hat 0}-(e_{\hat 0} \cdot \mathbb{S})\,\mathcal{A}^\mu\,,
\end{equation}
where $\mathbb{S}^\mu$ is a vector that is Fermi-Walker transported along  $e^{\mu}{}_{\hat 0}$.

\subsubsection{Spacetime Curvature as Measured by Static Observers}

Suppose that the static observer with orthonormal tetrad $e^{\mu}{}_{\hat \alpha}$ measures the spacetime curvature in our conformally flat  TGW spacetime. The components of the Riemann curvature tensor as measured by the observer are given by
\begin{equation}\label{L26}
R_{\hat \alpha \hat \beta \hat \gamma \hat \delta}(\tau) := R_{\mu \nu \rho \sigma}\,e^{\mu}{}_{\hat \alpha}\,
e^{\nu}{}_{\hat \beta}\,e^{\rho}{}_{\hat \gamma}\,e^{\sigma}{}_{\hat \delta}\,.
\end{equation}
These constitute the projection of the Riemann curvature tensor upon the tetrad frame of the static observer. Taking advantage of the symmetries of the Riemann tensor, this quantity can be represented by a $6\times6$ matrix $\mathcal{R} = (\mathcal{R}_{IJ})$, where the indices $I$ and $J$ range over the set $(01,02,03,23, 31,12)$. Thus we can write
\begin{equation}
\label{L27}
\mathcal{R}=\left[
\begin{array}{cc}
\mathcal{E} & \mathcal{B}\cr
\mathcal{B^{\dagger}} & \mathcal{S}\cr
\end{array}
\right]\,,
\end{equation}
where $\mathcal{E}$  and $\mathcal{S}$ are symmetric $3\times3$ matrices and  $\mathcal{B}$ is traceless. The tidal matrix $\mathcal{E}$ represents the ``electric" components of the curvature tensor as measured by the static observer, whereas $\mathcal{B}$ and $\mathcal{S}$ represent its ``magnetic" and ``spatial" components, respectively.  In the case under consideration,
$e^{\mu}{}_{\hat \alpha} = \Omega^{-1}\,\delta^{\mu}_{\alpha}$, so that Eq.~\eqref{L4} implies
\begin{equation}\label{L28}
\mathcal{E} = -\frac{1}{3} \Lambda \,\mathcal{I}\,, \qquad \mathcal{B} = 0\,,\qquad \mathcal{S} = \frac{1}{3} \Lambda\, \mathcal{I}\,,
\end{equation}
where $\mathcal{I}$ is the $3\times3$ identity matrix $\mathcal{I} =\,$diag$(1,1,1)$. These results, expected for anti-de Sitter spacetime, should be contrasted with the Weyl curvature of the Ricci-flat TGWs in Ref.~\cite{Bini:2018iyu}. The absence of the gravitomagnetic component of the Riemann tensor as measured by static fiducial observers is a peculiar feature of this propagating TGW that has no Weyl curvature. This point can be further illustrated via the Bel tensor in this case.

The super-energy-momentum tensor of a gravitational field is proportional to the symmetric and traceless quantity~\cite{Mashhoon:1996wa, Mashhoon:2003ax}
\begin{equation}\label{L28a}
\mathcal{T}_{\hat \alpha \hat \beta} = \bar{T}_{\mu \nu \rho \sigma}\,e^{\mu}{}_{\hat \alpha}\,
e^{\nu}{}_{\hat \beta}\,e^{\rho}{}_{\hat 0}\,e^{\sigma}{}_{\hat 0}\,,
\end{equation}
where $\bar{T}_{\mu \nu \rho \sigma}$ is the natural gravitational analog of the  energy-momentum tensor of the electromagnetic field
\begin{equation}\label{L28b}
 \bar{T}_{\mu \nu \rho \sigma} = \frac{1}{2} \,(R_{\mu \xi \rho \zeta}\,R_{\nu}{}^{\xi}{}_{\sigma}{}^{\zeta} + R_{\mu \xi \sigma \zeta}\,R_{\nu}{}^{\xi}{}_{\rho}{}^{\zeta}) -\frac{1}{4}\, g_{\mu \nu} \,
 R_{\alpha \beta \rho \gamma}\,R^{\alpha \beta}{}_{\sigma}{}^{\gamma}\,
\end{equation}
and was introduced by Bel in 1958~\cite{Bel}. In general, Bel's tensor, $\bar{T}_{\mu \nu \rho \sigma}$, is symmetric and traceless in its first pair of indices and symmetric in its second pair of indices.  In a Ricci-flat spacetime, Bel's tensor reduces to the completely symmetric and traceless Bel-Robinson tensor. 

For the conformally flat TGW with Riemann curvature~\eqref{L4}, the Bel tensor is given by
\begin{equation}\label{L28c}
 \bar{T}_{\mu \nu \rho \sigma} = \left(\frac{\Lambda}{3}\right)^2\,\Omega^4\,\left(\eta_{\mu \rho}\,\eta_{\nu \sigma} + \eta_{\mu \sigma}\,\eta_{\nu \rho} - \frac{1}{2}\,\eta_{\mu \nu}\,\eta_{\rho \sigma}\right)\,,
\end{equation}
which is traceless in its second pair of indices as well. In this case, the corresponding symmetric and traceless super-energy-momentum tensor as measured by the static fiducial observers is  proportional to
\begin{equation}\label{L28d}
\mathcal{T}_{\hat \alpha \hat \beta} = \left(\frac{\Lambda}{3}\right)^2\,\left(2\,\eta_{\hat \alpha \hat 0}\,\eta_{\hat \beta \hat 0} + \frac{1}{2}\,\eta_{\hat \alpha \hat \beta}\right)\,,
\end{equation}
which has the peculiar character of a perfect fluid \emph{at rest} with energy density $\Lambda^2/6$ and pressure $\Lambda^2/18$. The super-Poynting vector vanishes in this case in contrast to the Ricci-flat TGWs discussed in Ref.~\cite{Bini:2018iyu}. This circumstance illustrates the limitation of the super-momentum concept in the absence of Weyl curvature.

\subsubsection{Tidal Equations}

The equation for the motion of a timelike geodesic relative to our fiducial static observer within the framework of the Fermi coordinate system $(T, \mathbf{X})$ can be written as  
\begin{equation}\label{L29}
\frac{d^2X^{\hat i}}{dT^2}+ \mathcal{A}^{\hat i}+(\mathcal{E}^{\hat i}{}_{\hat j}+\mathcal{A}^{\hat i}\,\mathcal{A}_{\hat j})X^{\hat j} = 0\,,
\end{equation}
where the contribution of relative velocity has been neglected, see Appendix D. Here, $\mathcal{A}^{\hat 1} = - \varpi\,\cos\theta$, $\mathcal{A}^{\hat 2} = - \varpi\,\sin\theta$, $\mathcal{A}^{\hat 3} = s$ and  $\mathcal{E}_{\hat i \hat j} = R_{\hat 0 \hat i \hat 0 \hat j} = \varpi^2 \delta_{\hat i \hat j}$. Thus Eq.~\eqref{L29} can be expressed as 
\begin{equation}\label{L30}
\frac{d^2X^{\hat 1}}{dT^2}+ \varpi^2\,(1+\cos^2\theta)\,X^{\hat 1}+\varpi^2\,\sin\theta\,\cos\theta\,X^{\hat 2} - s\,\varpi\,\cos\theta\, X^{\hat 3} = \varpi\,\cos\theta\,,
\end{equation}
\begin{equation}\label{L31}
\frac{d^2X^{\hat 2}}{dT^2}+ \varpi^2\,\sin\theta\,\cos\theta\,X^{\hat 1} + \varpi^2\,(1+\sin^2\theta)\,X^{\hat 2} - s\,\varpi\,\sin\theta\, X^{\hat 3} = \varpi\,\sin\theta\,,
\end{equation}
\begin{equation}\label{L32}
\frac{d^2X^{\hat 3}}{dT^2} - s\,\varpi\,\cos\theta\, X^{\hat 1} - s\,\varpi\,\sin\theta\, X^{\hat 2}+ (s^2 + \varpi^2)\,X^{\hat 3}  = -s\,.
\end{equation}

It proves convenient to define $P_i$, $i = 1, 2, 3$,  as follows:
\begin{equation}\label{L33}
P_1 = \varpi\,\cos\theta\,X^{\hat 1} + \varpi\,\sin\theta\,X^{\hat 2} - s\,X^{\hat 3}\,, \qquad P_2 = - \sin\theta\, X^{\hat 1} + \cos\theta\, X^{\hat 2}\,,
\end{equation}
\begin{equation}\label{L34}
P_3 = s\,\cos\theta\,X^{\hat 1} + s\,\sin\theta\,X^{\hat 2} + \varpi\,X^{\hat 3}\,.
\end{equation}
Then, Eqs.~\eqref{L30}--\eqref{L32} can be written in terms of the new quantities as
\begin{equation}\label{L35}
\frac{d^2P_1}{dT^2}+ (s^2 + 2\, \varpi^2)\,P_1 = s^2 + \varpi^2\,,
\end{equation}
\begin{equation}\label{L36}
\frac{d^2P_2}{dT^2}+ \varpi^2\,P_2 = 0\,,\qquad   \frac{d^2P_3}{dT^2}+ \varpi^2\,P_3 = 0\,.
\end{equation}
 It is now straightforward to write down the general solution of the tidal equations in this case. That is, 
\begin{equation}\label{L37}
P_1 = \frac{s^2 + \varpi^2}{s^2 + 2\, \varpi^2}+ \xi_1\, \cos(\sqrt{s^2 + 2\,\varpi^2}\,T + \phi_1)\,,
\end{equation} 
\begin{equation}\label{L38}
P_2 = \xi_2\, \cos(\varpi\,T + \phi_2)\,, \qquad P_3 = \xi_3\, \cos(\varpi\,T + \phi_3)\,,
\end{equation}  
where $\xi_i$ and $\phi_i$, for $i = 1,2,3$, are integration constants.  Let us note that we can write
\begin{equation}\label{L39}
P_1 = \mathbb{D}_i\,X^{\hat i}\,, \qquad P_2 = \mathbb{N}_i\,X^{\hat i}\,, \qquad P_3 = \mathbb{E}_i\,X^{\hat i}\,,
\end{equation}   
where  $ \mathbb{D}_i$ are given by Eq.~\eqref{L16}, and $\mathbb{N}_i$ and $\mathbb{E}_i$ are defined here via Eqs.~\eqref{L33} and~\eqref{L34}.  That is, for $i = 1, 2, 3$, 
\begin{equation}\label{L40}
(\mathbb{D}_i) = (\varpi \,\cos\theta, \varpi \,\sin\theta, -s)\,, \quad (\mathbb{N}_i) = (- \sin\theta, \cos\theta, 0)\,, \quad (\mathbb{E}_i) = (s \,\cos\theta,  s\,\sin\theta, \varpi)\,,
\end{equation}
which are three spatially orthogonal vectors. It follows that
\begin{align}\label{L41}
X^{\hat i} =  {}& \frac{\mathbb{D}_i}{s^2 + 2\, \varpi^2} + \frac{\mathbb{D}_i}{s^2 + \varpi^2}\,\xi_1\, \cos(\sqrt{s^2 + 2\,\varpi^2}\,T + \phi_1) \\   \nonumber
{}& +\mathbb{N}_i\, \xi_2\, \cos(\varpi\,T + \phi_2) +  \frac{\mathbb{E}_i}{s^2 + \varpi^2}\, \xi_3\, \cos(\varpi\,T + \phi_3)\,.
\end{align}

\begin{figure}
\includegraphics[scale=0.45]{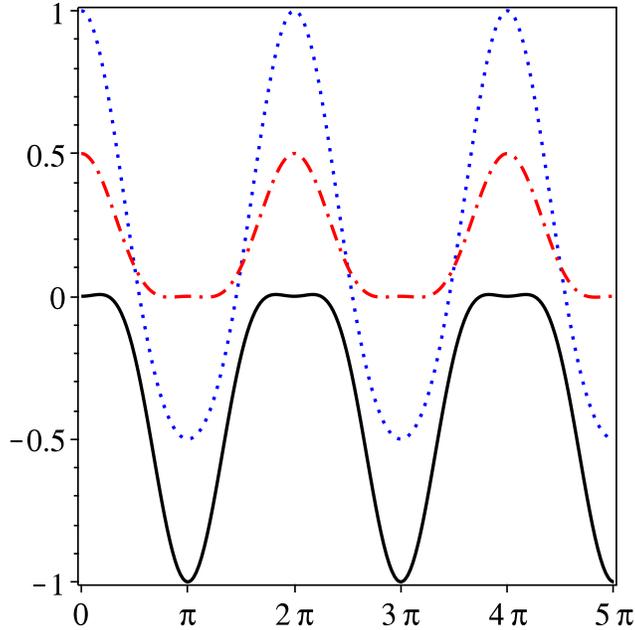}
\caption{\label{figV} Plot of $\varpi\,X^{\hat i}$ versus $\varpi\,T$ for $i=1$ (blue dot ), $i=2$ (red dash-dot) 
and $i=3$ (solid black),  where $X^{\hat i}$  is given by Eq.~\eqref{L41}. Initial conditions at $T = 0$ are chosen such that $(X^{\hat 1}, X^{\hat 2}, X^{\hat 3}) = (1, 0.5, 0)$ and $dX^{\hat i}/dT=0$  for $i = 1,2,3$. Thus, $\phi_1 = \phi_2 = \phi_3 = 0$. Moreover, $s = \sqrt{2}\,\varpi$ and $\theta = 45^{\circ}$, so that $\xi_1 = 3(\sqrt{2} - 1)/4$, $\xi_2 = -\sqrt{2}/(4\,\varpi)$ and $\xi_3 = 3/2$.}
\end{figure}

The \emph{transverse} character of linearized gravitational waves in GR is well known. Twisted gravitational waves that are Ricci-flat exhibit in addition a longitudinal component as well~\cite{Bini:2018gbq}. For a general discussion of the corresponding longitudinal component in the presence of Weyl curvature tensor, see Refs.~\cite{PZ, Podolsky:2012he}.  However, the Weyl conformal curvature tensor vanishes for our special solution; for a general discussion of the Jacobi equation in this case, see Ref.~\cite{CH}. The longitudinal component in the absence of Weyl curvature is given by $X^{\hat 3}$ in the present case, which is along the direction of wave propagation and can be obtained from Eq.~\eqref{L41} for $i = 3$.  This longitudinal feature is illustrated in Figure~\ref{figV}.

\section{Discussion}

Three classes of TGW solutions in the presence of a cosmological constant have been presented in Sections II--IV. These are generally implicit, as each GR solution depends upon the solution of an ordinary differential equation for $A(w)$, where $w$ depends linearly on spacetime coordinates $(t, x, y, z)$. Nevertheless, we have determined the general behavior of these  solutions and their curvature singularities; moreover, we have characterized the algebraically special nature of their Weyl curvature tensors. Among the new solutions, there is a simple unique conformally flat solution with $g_{\mu \nu} = \Omega^2\,\eta_{\mu \nu}$ that has a conformal factor $\Omega$ given by  $\Omega^{-1} = s\,u + p\,x + q\,y$, where $(s, p, q)$ are constants subject to $p^2+q^2 = -\Lambda/3$. The wave front for this simple TGW has constant \emph{negative} Gaussian curvature determined by the cosmological constant $\Lambda$, namely, $K_G = \Lambda/3$. This special  explicit solution for negative cosmological constant represents part of anti-de Sitter spacetime and has been studied in detail in the previous section. That is, the timelike geodesics of this solution have been worked out and the deviation of these geodesics relative to the world lines of static observers in this spacetime have been examined in connection with measurements of static fiducial observers. These observers in anti-de Sitter spacetime have $(t, x, y, z)$ coordinates adapted to their motions such that the constant negative curvature AdS spacetime appears in the form of a TGW in these adapted coordinates. 

All of the known TGWs, regardless of the presence of the cosmological constant, have wave fronts with \emph{negative} Gaussian curvature. It is not known whether this is a general feature of TGWs or occurs due to the formal simplicity of the solutions that have been found thus far.

\section*{ACKNOWLEDGMENTS}

B.\,M. is grateful to Donato Bini for his valuable comments on the manuscript.

\appendix

\section{Gaussian Curvature of the Wave Front}

Consider a TGW spacetime with metric of the form
\begin{equation}\label{A1}
ds^2 = -\gamma_0\,dt^2 + \gamma_1\,dx^2+ 2\gamma_2\, dx\,dy + \gamma_3\,dy^2 + \gamma_0\,dz^2\,,
\end{equation}
where $\gamma_\mu = \gamma_\mu (u, x, y)$ and $u := t-z$. The metric of the wave front is given by 
\begin{equation}\label{A2}
d\sigma^2  = \gamma_1(u, x, y)\,dx^2+ 2\gamma_2(u, x, y)\, dx\,dy + \gamma_3(u, x, y)\,dy^2\,,
\end{equation}
where $u$ is a constant in this case. It is possible to show that for any $(t, x, y, z)$, 
\begin{equation}\label{A3}
^{(s)}{}R_{xyxy}~=~^{(\sigma)}{}R_{xyxy}\,,
\end{equation}
where  $^{(s)}R_{xyxy}$ is a component of the Riemann curvature tensor for metric~\eqref{A1}, while $^{(\sigma)}R_{xyxy}$ is the corresponding component for metric~\eqref{A2}.

For metric~\eqref{A2}, the Gaussian curvature $K_G$ is given by~\cite{R1} 
\begin{equation}\label{A4}
K_G = \frac{^{(\sigma)}R_{xyxy}}{\Delta}\,, \qquad \Delta = \gamma_1\,\gamma_3 - \gamma_2^2 > 0\,.
\end{equation}
From
\begin{equation}\label{A5}
\gamma_3\, ^{(\sigma)}R_{xyxy} = \Delta\, ^{(\sigma)}R^{x}{}_{yxy}\,,
\end{equation}
we get the simple relation
\begin{equation}\label{A6}
K_G = \frac{1}{\gamma_3}\,^{(\sigma)}R^{x}{}_{yxy}\,.
\end{equation}
Using the standard formula for the Riemann tensor, we find
\begin{align} \label{A7}
K_G = {} & \frac{1}{2\,\Delta}\,(2\,\gamma_{2,xy} -\gamma_{1,yy} - \gamma_{3,xx}) -  \frac{\gamma_3}{4\,\Delta^2}\,(2\,\gamma_{1,x}\,\gamma_{2,y} -\gamma_{1,x}\,\gamma_{3,x}  - \gamma_{1,y}^2) \\ 
\nonumber  & -  \frac{\gamma_1}{4\,\Delta^2}\,(2\,\gamma_{2,x}\,\gamma_{3,y} -\gamma_{1,y}\,\gamma_{3,y} - \gamma_{3,x}^2) +  \frac{\gamma_2}{4\,\Delta^2}\,[\gamma_{1,x}\,\gamma_{3,y} - 2\,\gamma_{1,y}\,\gamma_{3,x} \\
\nonumber & + (2\,\gamma_{2,x} -\gamma_{1,y})\, (2\gamma_{2,y} - \gamma_{3,x})]\,, 
\end{align}
where a comma denotes partial differentiation.

In the special case where $\gamma_2 = 0$, $\gamma_1 = e^{B}$ and $\gamma_3 = e^{C}$, Eq.~\eqref{A7} reduces to formula (B2) of Appendix B of Ref.~\cite{Bini:2018gbq}.

\section{Gravitational Field Equations}

The purpose of this appendix is to present the gravitational field equations~\eqref{I12} for metric~\eqref{I5} when condition~\eqref{I13} is satisfied.  It follows from the admissibility conditions for the coordinates that the metric can be written as 
\begin{equation}\label{B1}
ds^2 = - e^A\, du\,dv+ e^B\,dx^2+ 2\, h\, dx\,dy + e^C\,dy^2\,,
\end{equation} 
where $A, B, C$ and $h$ are functions of $w = s\,u + p\,x + q\,y$ and $f(w) := \exp(B+C) - h^2 > 0$. The metric is invariant under the exchange of $(x, B, p)$ with $(y, C, q)$, respectively. This invariance is then reflected in the field equations $R_{\mu \nu } = \Lambda\,g_{\mu \nu}$. In this connection, it is convenient to introduce   
\begin{equation}\label{B2}
H_A = h' - hA'\,, \qquad H_B = h' - hB'\,, \qquad H_C = h' - hC'\,, 
\end{equation} 
where $h' := dh/dw$, etc. The homogeneous field equations are then given by $R_{uu} = 0$, $R_{ux} = 0$, $R_{uy} = 0$ and $R_{xy} = 0$, which can be expressed as
\begin{align}\label{B3}
{} &(f+h^2) (H_B + H_C)^2 - 2\,f\,H_B\,H_C - 2\,fh[A'(H_B + H_C) - (H'_B + H'_C)] \\ \nonumber 
{} &-fh^2(B'-C')^2 - f^2\,[2B'' + B'^2 + 2C'' + C'^2 -2A'(B' +C')] = 0\,, 
\end{align} 
\begin{align}\label{B4}
{} & q\,e^B\,[h (H_B + H_C) H_B + 2\,fH'_B + f(B'-C')H_B] + p\,[ h^2(H_B + H_C)H_C  \\ \nonumber 
{} & -fh(A'H_B+A'H_C +B'H_C +C'H_B -2\,H'_C) \\  \nonumber
{} &- f^2\,(2C'' + C'^2 + 2A'' -A' B' -A'C' -B'C')] = 0\,, 
\end{align} 
\begin{align}\label{B5}
{} & p\,e^C\,[h (H_B + H_C) H_C + 2\,fH'_C - f(B'-C')H_C] + q\,[ h^2(H_B + H_C)H_B  \\ \nonumber 
{} & -fh(A'H_B+A'H_C +B'H_C +C'H_B -2\,H'_B) \\  \nonumber
{} &- f^2\,(2B'' + B'^2 + 2A'' -A' B' -A'C' -B'C')] = 0\, 
\end{align} 
and
\begin{align}\label{B6}
{} & h\,q^2\,e^B\,[h (H_B + H_C) B' + f\,(2\,B'' + B'^2 +2\, A'B' - B'C')]  \\ \nonumber 
{} & + h\,p^2\,e^C\,[h (H_B + H_C) C' + f\,(2\,C'' + C'^2 +2\, A'C' - B'C')]  \\  \nonumber
{} & +2\,pq\,[- 2fhh'' - h^2 h' (H_B + H_C) + fh(B' + C')H_A  \\ \nonumber 
{} &+ f^2\,(2A'' + A'^2 -A' B' -A'C')] = 0\,, 
 \end{align} 
respectively. Furthermore, the three inhomogeneous field equations, namely, $R_{uv} = -(\Lambda/2)\,\exp(A)$, $R_{xx} = \Lambda\,\exp(B)$ and $R_{yy} = \Lambda\,\exp(C)$ are given by
\begin{align}\label{B7}
- 4\,f^2 \Lambda = {} & p^2\,e^C\,[h (H_B + H_C) A' + f\,(2\,A'' + 2\,A'^2+ A'C' - A'B')]  \\  \nonumber
{} &  + q^2\,e^B\,[h (H_B + H_C) A' + f\,(2\,A'' + 2\,A'^2+ A'B' -A'C')]  \\ \nonumber 
{} & - 2\,pq\,[h^2 (H_B + H_C) A' + 2\,fh'A' +fh(2\,A'' + 2\,A'^2 - A'B' -A'C')]\,,
\end{align} 
\begin{align}\label{B8}
- 4\,f^2 \Lambda = {} & q^2\,e^B\,[h (H_B + H_C) B' + f\,(2\,B'' + B'^2+ 2\,A'B' -B'C')]  \\  \nonumber
{} & + p^2\,e^{-B}\,[h^3 (H_B + H_C) C' +2\, fh(hC'' + h'C'+ 2\,h'A' - h A'B' - hB'C')  \\ \nonumber 
{} &+ f^2\,(4A'' + 2 A'^2 + 2C'' + C'^2 - 2A' B'  -B'C')]  \\ \nonumber
{} & - 2\,pq\,[h h' (H_B + H_C) + 2\,fh'' +fh'(2\,A' - B' - C')]\,
 \end{align}  
and
\begin{align}\label{B9}
- 4\,f^2 \Lambda = {} & p^2\,e^C\,[h (H_B + H_C) C' + f\,(2\,C'' + C'^2+ 2\,A'C' -B'C')]  \\  \nonumber
{} & + q^2\,e^{-C}\,[h^3 (H_B + H_C) B' +2\, fh(hB'' + h'B'+ 2\,h'A' - h A'C' - hB'C')  \\ \nonumber 
{} &+ f^2\,(4A'' + 2 A'^2 + 2B'' + B'^2 - 2A' C'  -B'C')]  \\ \nonumber
{} & - 2\,pq\,[h h' (H_B + H_C) + 2\,fh'' +fh'(2\,A' - B' - C')]\,.
 \end{align} 

The field equations employed in this paper can be obtained as special cases of the results given in this appendix.

\section{Conformally Flat TGW Solution of $R_{\mu \nu } = \Lambda\,g_{\mu \nu}$}

We look for solutions of the gravitational field equations  $R_{\mu \nu } = \Lambda\,g_{\mu \nu}$ with a conformally flat TGW metric of the form
\begin{equation}\label{C1}
ds^2 = e^{A(u, x, y)} (-du\,dv + dx^2 + dy^2)\,.
\end{equation}
The corresponding field equations consist of three equations with source $\Lambda$, namely,
\begin{equation}\label{C2}
-2 R_{uv} = R_{xx} = R_{yy} = \Lambda\,e^{A(u, x, y)}\,,
\end{equation}
which can be written out explicitly for metric~\eqref{C1} as
\begin{equation}\label{C3}
A_{,xx} + A_{,yy} + (A_{,x})^2 + (A_{,y})^2 = -2\,\Lambda\,e^{A}\,,
\end{equation}
\begin{equation}\label{C4}
3\,A_{,xx} + A_{,yy} + (A_{,y})^2 = -2\,\Lambda\,e^{A}\,,
\end{equation}
\begin{equation}\label{C5}
3\,A_{,yy} + A_{,xx} + (A_{,x})^2 = -2\,\Lambda\,e^{A}\,,
\end{equation}
respectively. Furthermore, there are four nontrivial source-free field equations
\begin{equation}\label{C6}
R_{uu} = R_{ux} = R_{uy} = R_{xy} = 0\,,
\end{equation}
which can be expressed as 
\begin{equation}\label{C7}
2\,A_{,uu} =  (A_{,u})^2\,,
\end{equation}
\begin{equation}\label{C8}
2\,A_{,ux} =  A_{,u}\,A_{,x}\,, \qquad 2\,A_{,uy} =  A_{,u}\,A_{,y}\,, \qquad  2\,A_{,xy} =  A_{,x}\,A_{,y}\,,
\end{equation}
respectively. 

Let us subtract Eq.~\eqref{C3} from Eqs.~\eqref{C4} and~\eqref{C5} to get 
\begin{equation}\label{C9}
2\,A_{,xx} =  (A_{,x})^2\,, \qquad 2\,A_{,yy} =  (A_{,y})^2\,.
\end{equation}
Then, Eq.~\eqref{C3} reduces to 
\begin{equation}\label{C10}
 (A_{,x})^2 + (A_{,y})^2 = -\frac{4}{3}\,\Lambda\,e^{A}\,.
\end{equation}
Next, by virtue of Eq.~\eqref{C7} we have 
\begin{equation}\label{C11}
\left(e^{-\frac{1}{2}\,A}\right)_{,uu} = \frac{1}{4}\,[2\,A_{,uu} -  (A_{,u})^2]\,e^{-\frac{1}{2}\,A} = 0\,,
\end{equation}
which implies that $\exp{(-A/2)}$ is a \emph{linear} function of $u$. Similarly, it follows from Eq.~\eqref{C9} that $\exp{(-A/2)}$ depends linearly upon $x$ and $y$ as well. Thus, we can write
\begin{equation}\label{C12}
e^{-\frac{1}{2}\,A} = s\,u + p\,x + q\,y\,
\end{equation}
plus a constant that can always be removed by a simple coordinate translation. The integration constant $s$ is arbitrary, while Eq.~\eqref{C10} implies
\begin{equation}\label{C13}
 p^2 + q^2 = -\frac{1}{3}\,\Lambda\,.
\end{equation}
The remaining field Eqs.~\eqref{C8} are all satisfied by this unique class of conformally flat solutions with constant parameters $(s, p, q)$ subject to restriction~\eqref{C13}. This gravitational field disappears in the absence of a negative cosmological constant.

\section{Deviation Equation in Fermi Coordinates}

Consider an arbitrary static observer with proper time $\tau$ following a world line $\bar{x}^\mu(\tau)$. Let $e^{\mu}{}_{\hat \alpha}(\tau)$ be a Fermi-Walker transported tetrad along $\bar{x}^\mu(\tau)$. At each event on the observer's path, we imagine the set of all spacelike geodesics that are orthogonal to the world line at $\bar{x}^\mu(\tau)$ and form a spacelike hypersurface. Let 
$x^\mu$ be an event on this hypersurface that can be connected to $\bar{x}^\mu(\tau)$ with a \emph{unique} spacelike geodesic of proper length $\varsigma$. We assign to event $x^\mu$ Fermi coordinates $X^{\hat \mu} = (T, X^{\hat i})$, where
\begin{equation}\label{D1}
T := \tau\,, \qquad X^{\hat i} := \varsigma\, \sigma^\mu(\tau)\, e_{\mu}{}^{\hat i}(\tau)\,.
\end{equation}
The \emph{unit} spacelike vector tangent  at $\bar{x}^\mu(\tau)$ to the unique spacelike geodesic connecting $\bar{x}^\mu(\tau)$ with $x^\mu$ is denoted by $\sigma^\mu$; hence, $\sigma^\mu(\tau)\,e_{\mu \,\hat 0}(\tau) = 0$. It is clear that the reference observer occupies the spatial origin of the Fermi coordinate system.

We wish to study the timelike geodesic equation in the Fermi coordinate system. In this way, we can determine the motion of a free test particle relative to the fiducial static observer.  Neglecting the relative velocity, the reduced geodesic equation can be expressed as
\begin{equation}\label{D2}
\frac{d^2X^{\hat i}}{dT^2}+\mathcal{A}^{\hat i}(T)+[\mathcal{E}^{\hat i}{}_{\hat j}(T)+\mathcal{A}^{\hat i}(T)\,\mathcal{A}_{\hat j}(T)]\,X^{\hat j} = 0\,,
\end{equation}
where $\mathcal{A}^{\hat i}$ is the 4-acceleration of the fiducial static observer projected upon its frame and $\mathcal{E}_{\hat i \hat j}$ are the corresponding components of the tidal matrix, namely, 
\begin{equation}\label{D3}
\mathcal{A}^{\hat i}(T) = \frac{De^{\mu}{}_{\hat 0}}{d\tau}\,e_{\mu}{}^{\hat i} = \mathcal{A}^\mu\,e_{\mu}{}^{\hat i}\,, \qquad \mathcal{E}_{\hat i \hat j}(T) = R_{\hat 0 \hat i \hat 0 \hat j} = R_{\mu \nu \rho \sigma}\,e^{\mu}{}_{\hat 0}\,
e^{\nu}{}_{\hat i}\,e^{\rho}{}_{\hat 0}\,e^{\sigma}{}_{\hat j}\,.
\end{equation}   
For background material on the equations of motion in Fermi coordinates, we refer to Refs.~\cite{Bini:2017uax, mas77, CM3} and the references cited therein. The Fermi coordinate system is generally admissible in a certain cylindrical spacetime domain around $\bar{x}^\mu(\tau)$.


\begin{thebibliography}{00}


\bibitem{Bini:2018gbq} 
  D.~Bini, C.~Chicone and B.~Mashhoon,
  ``Twisted Gravitational Waves'',
  Phys.\ Rev.\ D {\bf 97}, no. 6, 064022 (2018)
  [arXiv:1801.06003 [gr-qc]].
  
\bibitem{Bini:2018iyu} 
  D.~Bini, C.~Chicone, B.~Mashhoon and K.~Rosquist,
  ``Spinning Particles in Twisted Gravitational Wave Spacetimes'',
  Phys.\ Rev.\ D {\bf 98}, no. 2, 024043 (2018)
  [arXiv:1805.07080 [gr-qc]].



\bibitem{Rosquist:2018ore} 
  K.~Rosquist, D.~Bini and B.~Mashhoon,
  ``Twisted Gravitational Waves of Petrov Type \emph{D}'',
  Phys.\ Rev.\ D {\bf 98}, no. 6, 064039 (2018)
  [arXiv:1807.09214 [gr-qc]].
  
\bibitem{Brink}
H.~W.~Brinkmann, ``Einstein spaces which are mapped conformally on each other", 
Math. Ann. \textbf{94}, 119-145 (1925).




 \bibitem{BPR} 
H.~Bondi, F.~A.~E.~Pirani and I.~Robinson,
 ``Gravitational waves in general relativity III. Exact plane waves",
Proc. R. Soc. A \textbf{251}, 519 (1959).

\bibitem{Khan:1971vh} 
  K.~A.~Khan and R.~Penrose,
  ``Scattering of two impulsive gravitational plane waves'',
  Nature {\bf 229}, 185-186 (1971).

\bibitem{JBG} 
J.~B.~Griffiths,
\emph{Colliding Plane Waves in General Relativity}
(Oxford University Press,  Oxford, England, 1991).

\bibitem{R1}
H.~Stephani, D.~Kramer, M.~MacCallum, C.~Hoenselaers and E.~Herlt, 
\emph{Exact Solutions of Einstein's Field Equations}, 2nd ed. (Cambridge University Press, Cambridge, England,  2003).


\bibitem{Griffiths:2009dfa} 
  J.~B.~Griffiths and J.~Podolsk\'y,
  \emph{Exact Space-Times in Einstein's General Relativity} (Cambridge University Press, Cambridge, England,  2009).
  
\bibitem{Bini:2012ht} 
  D.~Bini, C.~Chicone and B.~Mashhoon,
  ``Spacetime Splitting, Admissible Coordinates and Causality'',
  Phys.\ Rev.\ D {\bf 85}, 104020 (2012)
  [arXiv:1203.3454 [gr-qc]].


\bibitem{Bini}
D.~Bini  (private communication to BM, 2018). 


\bibitem{Skea}
J.~E.~F.~Skea, 
``The invariant classification of conformally flat pure radiation spacetimes",
 Classical Quantum Gravity  {\bf 14}, 2393 (1997).



\bibitem{Griffiths:1998sp} 
  J.~B.~Griffiths and J.~Podolsk\'y,
  ``Interpreting a conformally flat pure radiation space-time'',
  Classical Quantum Gravity  {\bf 15}, 3863 (1998)
  [arXiv: gr-qc/9808061].

\bibitem{Chicone:2010xr} 
  C.~Chicone, B.~Mashhoon and K.~Rosquist,
  ``Cosmic Jets'',
  Phys.\ Lett.\ A {\bf 375}, 1427-1430 (2011)
  [arXiv:1011.3477 [gr-qc]].
  
  
 \bibitem{Bini:2014esa} 
  D.~Bini and B.~Mashhoon,
  ``Peculiar velocities in dynamic spacetimes'',
  Phys.\ Rev.\ D {\bf 90},  024030 (2014)
  [arXiv:1405.4430 [gr-qc]].
  
  
  
  
\bibitem{Bini:2017qnd} 
  D.~Bini, C.~Chicone and B.~Mashhoon,
  ``Anisotropic gravitational collapse and cosmic jets'',
  Phys.\ Rev.\ D {\bf 96}, 084034 (2017)
  [arXiv:1708.01040 [gr-qc]].

  
\bibitem{HE}
S.~W.~Hawking and G.~F.~R.~Ellis,
\emph{The Large Scale Structure of Space-Time} 
(Cambridge University Press, Cambridge, England, 1973).
  

  
\bibitem{Mashhoon:1996wa} 
  B.~Mashhoon, J.~C.~McClune and H.~Quevedo,
  ``Gravitational superenergy tensor'',
  Phys.\ Lett.\ A {\bf 231}, 47 (1997)
  [arXiv: gr-qc/9609018].

  
\bibitem{Mashhoon:2003ax} 
  B.~Mashhoon,
  ``Gravitoelectromagnetism: A brief review,''
 in: \emph{The Measurement of Gravitomagnetism:
A Challenging Enterprise}, edited by L. Iorio, Ch. 3, pp. 29-39 (NOVA Science, Hauppage,
New York, 2007)
 [arXiv: gr-qc/0311030].
  
\bibitem{Bel}
L.~Bel, 
``Sur la radiation gravitationnelle",
Compt. Rend. {\bf 247}, 1094-1096 (1958).
  

\bibitem{PZ}
P.~Szekeres, ``The gravitational compass", J. Math. Phys. (N.Y.) {\bf 6}, 1387 (1965).

\bibitem{Podolsky:2012he} 
  J.~Podolsk\'y and R.~\u Svarc,
  ``Interpreting spacetimes of any dimension using geodesic deviation",
  Phys.\ Rev.\ D {\bf 85}, 044057 (2012)
  [arXiv:1201.4790 [gr-qc]].

\bibitem{CH}
R.~F.~Crade and G.~S.~Hall,
``The deviation of timelike geodesics in space-time",
Phys. Lett. A {\bf 85}, 313-315 (1981). 
  
\bibitem{Bini:2017uax} 
  D.~Bini, C.~Chicone and B.~Mashhoon,
  ``Relativistic tidal acceleration of astrophysical jets'',
  Phys.\ Rev.\ D {\bf 95}, 104029 (2017)
  [arXiv:1703.10843 [gr-qc]].
  

\bibitem{mas77}
B. Mashhoon,
``Tidal radiation'',
Astrophys. J.\ {\bf 216}, 591-609 (1977).


\bibitem{CM3}
C.~Chicone and B.~Mashhoon, ``Explicit Fermi coordinates and tidal dynamics in de Sitter and G\"odel spacetimes", Phys. Rev. D {\bf 74}, 064019 (2006).





  


\end{thebibliography}
\end{document}